\documentclass[twocolumn, times, trackchanges]{aastex63_bren}
\usepackage{CJK}
\usepackage{booktabs} 
\usepackage{xcolor, soul}
\usepackage{amsmath, amssymb, bm}

\newcommand{\WDmodelsurl}{\href{https://github.com/SihaoCheng/WD_models}{\url{https://github.com/SihaoCheng/WD_models}}}

\newcommand{\wdurl}{\href{http://www.astro.umontreal.ca/~bergeron/CoolingModels/}{\url{http://www.astro.umontreal.ca/~bergeron/CoolingModels/}}}

\newcommand{\schengurl}{\href{https://pages.jh.edu/~scheng40/Q_branch}{\url{https://pages.jh.edu/~scheng40/Qbranch}}}

\newcommand{\MWDDurl}{\href{http://www.montrealwhitedwarfdatabase.org}{\url{http://www.montrealwhitedwarfdatabase.org}}}

\newcommand{\vL}{v_{\rm L}}
\newcommand{\vB}{v_{\rm B}}
\newcommand{\mathbfit}{\bm}
\newcommand{\bvT}{\mathbfit{v}_{\rm T}}
\newcommand{\bvzero}{\mathbfit{v}_0}

\begin{document}
\begin{CJK*}{UTF8}{gbsn}
\received{2019 June 6}
\revised{2019 September 1}
\accepted{2019 September 4}
\submitjournal{\apj}

\author[0000-0002-9156-7461]{Sihao Cheng (程思浩)}
\affiliation{Department of Physics and Astronomy, The Johns Hopkins University, 3400 N Charles Street, Baltimore, MD 21218, USA}

\author[0000-0001-7453-9947]{Jeffrey D. Cummings}
\affiliation{Department of Physics and Astronomy, The Johns Hopkins University, 3400 N Charles Street, Baltimore, MD 21218, USA}

\author{Brice M\'enard}
\affiliation{Department of Physics and Astronomy, The Johns Hopkins University, 3400 N Charles Street, Baltimore, MD 21218, USA}
\affiliation{Kavli Institute for the Physics and Mathematics of the Universe, University of Tokyo, Kashiwa 277-8583, Japan}

\title{A Cooling Anomaly of High-Mass White Dwarfs}
\shorttitle{A Cooling Anomaly of White Dwarfs}

\shortauthors{Cheng, Cummings, \& M\'enard}
\email{s.cheng@jhu.edu}

\begin{abstract}
Recently, the power of {\it Gaia} data has revealed an enhancement of high-mass white dwarfs (WDs) on the Hertzsprung--Russell diagram, called the Q branch. This branch is located at the high-mass end of the recently identified crystallization branch. Investigating its properties, we find that the number density and velocity distribution on the Q branch cannot be explained by the cooling delay of crystallization alone, suggesting the existence of an extra cooling delay. To quantify this delay, we statistically compare two age indicators -- the dynamical age inferred from transverse velocity, and the photometric isochrone age -- for more than one thousand high-mass WDs (1.08--1.23~$M_\odot$) selected from {\it Gaia} Data Release 2. We show that about $6\,\%$ of the high-mass WDs must experience an 8 Gyr extra cooling delay on the Q branch, in addition to the crystallization and merger delays. This cooling anomaly is a challenge for WD cooling models. We point out that $\rm^{22}Ne$ settling in C/O-core WDs could account for this extra cooling delay.\\
\end{abstract}


\section{Introduction}
\label{sec:introduction}

Until recently, explorations of the white dwarf region in the Hertzsprung--Russell (H--R) diagram were severely limited by the number of objects with available distance estimates. The European Space Agency {\it Gaia} mission \citep{GaiaCollaboration_2016} has changed this situation drastically. {\it Gaia} is an all-sky survey of astrometry and photometry for stars down to 20.7 magnitude. The H--R diagram of white dwarfs generated by {\it Gaia} Data Release 2 (DR2) reveals three branch-like features, called the A, B, and Q branches\footnote{Named after the presence of DA, DB, and DQ white dwarfs \citep{GaiaCollaboration_2018a}, respectively. DA white dwarfs have hydrogen lines in their spectr, DB and DQ white dwarfs have helium and carbon lines, respectively.} in Figure 13 of \citet{GaiaCollaboration_2018a}. The A and B branches have been understood as standard-mass white dwarfs ($m_\text{WD}\sim 0.6\,M_\odot$) with hydrogen-rich and helium-rich atmospheres, respectively \citep[e.g.,][]{Bergeron_2019}. However, the Q branch, as an enhancement of high-mass white dwarfs ($m_\text{WD}>1.0\,M_\odot$), is still not fully understood. This is a challenge to current white dwarf evolutionary models and an opportunity for studying high-mass white dwarfs. 
    
On the H--R diagram, white dwarfs evolve along their cooling tracks. Unlike the A and B branches, the Q branch is not aligned with any cooling track or isochrone, suggesting that it is caused by a delay of cooling instead of a peak in mass or age distribution. This cooling delay makes white dwarfs pile up on the Q branch. 
The Q branch coincides with the high-mass region of the crystallization branch identified by \citet{Tremblay_2019}. As a liquid-to-solid phase transition in the white dwarf core, crystallization releases energy through latent heat \citep[e.g.,][]{vanHorn_1968} and phase separation \citep[e.g.,][]{Garcia-Berro_1988, Segretain_1994, Isern_1997}, which can indeed create a cooling delay.
However, the observed pile-up on the Q branch is higher and narrower than expected from the standard crystallization models \citep[][figure 4]{Tremblay_2019}, suggesting that an extra cooling delay may exist in addition to crystallization. 

In this paper, we investigate this cooling anomaly using kinematic information of high-mass white dwarfs in {\it Gaia} DR2. In Section~\ref{sec:data} we describe our white dwarf sample; in Section~\ref{sec:extra cooling delay} we show strong evidence for the existence of an extra cooling delay on the Q branch; in Section~\ref{sec:quantitative inference} we build a model for the white dwarf velocity distribution and use our {\it Gaia} sample to constrain the properties of this cooling anomaly; in Section~\ref{sec:results} we present the best-fit values of these properties and as a byproduct of our model of our analysis, the fraction of double-WD merger products among high-mass white dwarfs; in Section~\ref{sec:22Ne} we show that $^{22}$Ne settling in massive C/O-core white dwarfs is a promising physical origin of this extra cooling delay; in Section~\ref{sec:discussion} we examine other aspects of the Q brancn, which provide evidence that the extra delayed white dwarfs are also double-WD merger products; and in Section~\ref{sec:conclusion} we conclude on our findings.

\begin{figure*}
    \centering
    \includegraphics[width=0.8\textwidth]{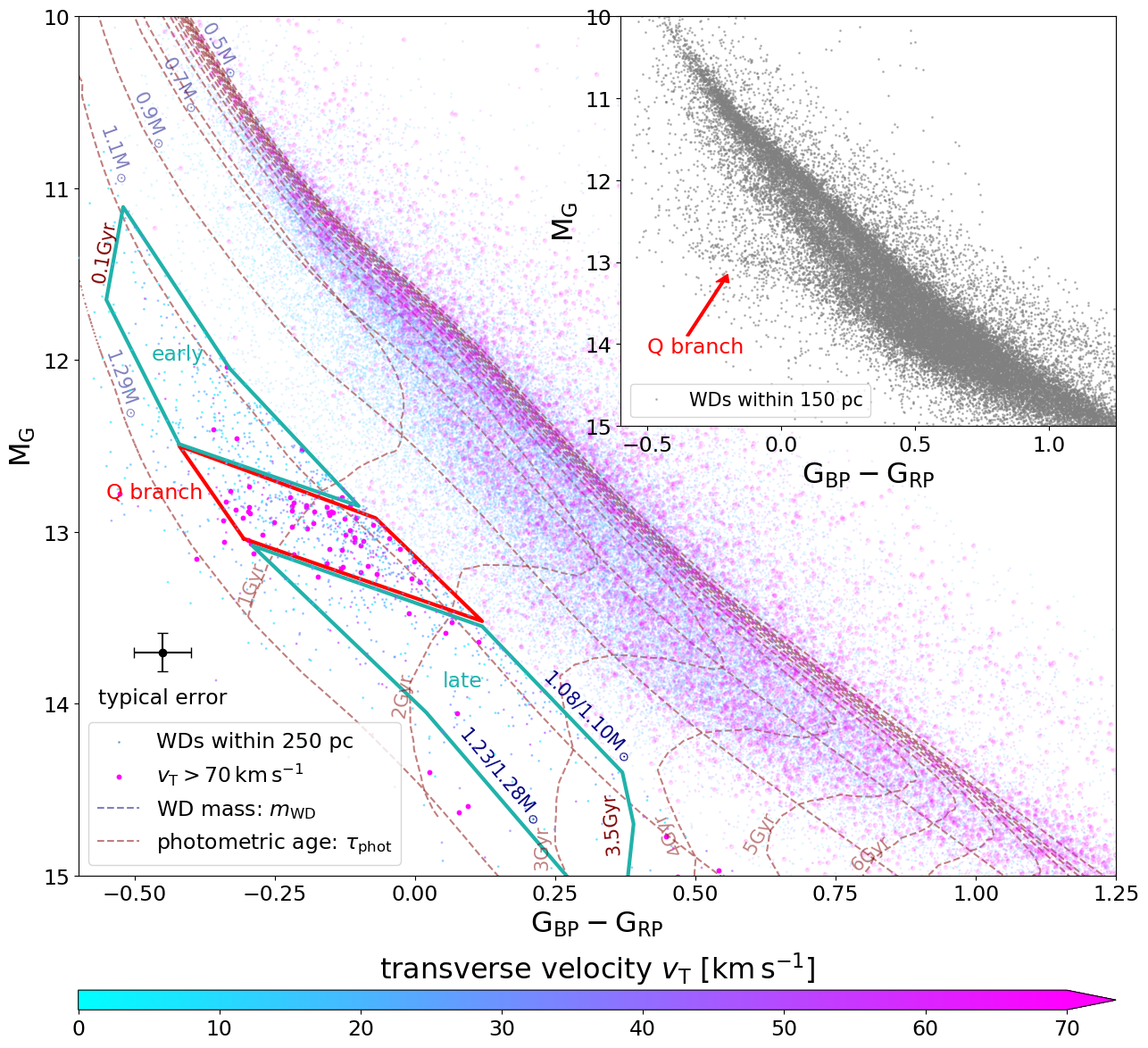}
    \caption{The H--R diagram of WDs selected from Section~\ref{sec:quality cuts}. In the top right panel we use the 150 pc  sample to show the number-density  distribution with a higher contrast. The Q branch is marked by the red arrow. 
    In the main panel, we show our main  250 pc sample color-coded with transverse velocities $v_{\rm T}$. Fast WDs ($v_{\rm T}>$ 70 km s$^{-1}$) are emphasized by large symbols, and high-mass WDs ($>1.08\,M_\odot$) are emphasized by high symbol opacity.
    The grid of WD mass and photometric age is also plotted (using the O/Ne model for high-mass WDs). For the mass range marked by dark blue texts, the first (second) number corresponds to the O/Ne (C/O) model.
    \label{fig:WD_HR_70}}
\end{figure*}

\section{Data}
\label{sec:data}

We use data from {\it Gaia} DR2 \citep{GaiaCollaboration_2018b}, which for the first time provides parallaxes $\varpi$ and proper motions $\mu$ that are derived purely from {\it Gaia} measurements \citep{Lindegren_2018}. {\it Gaia} DR2 also provides Vega magnitudes of three wide passbands \citep{Riello_2018, Evans_2018}: the $G$ band spans from 350 to 1000~nm, and the $G_{\rm BP}$ and $G_{\rm BP}$ bands are mainly the blue and red parts of the $G$ band, separated at the H$\alpha$ transition \citep{GaiaCollaboration_2016}.

\subsection{Quality cuts}
\label{sec:quality cuts}

\citet{GentileFusillo_2019} have compiled a catalog of {\it Gaia} DR2 white dwarfs based on the $G$-band absolute magnitude, {\it Gaia} color index, and some quality cuts. To select white dwarfs with high-precision measurements, we further apply the following quality cuts:
\begin{align}
    \sigma_{ G_{\rm BP}-G_{\rm RP}}&<0.10\,,\\
    \sigma_{\mu}/\varpi&<\text{2 km s}^{-1}\,,\\
    \texttt{parallax\_over\_error }& > 8\,,\\
    \texttt{astrometric\_excess\_noise }& < 1.5\,,\\
    \texttt{phot\_bp\_rp\_excess\_factor }& < 1.4\,,\\
    1/\varpi &< \text{250 pc}\,,
\end{align}
where the color error $\sigma_{G_{\rm BP}-G_{\rm RP}}$ is the combined photometric errors in $G_{\rm BP}$ and $G_{\rm RP}$ bands, the proper motion error $\sigma_{\mu}$ is the combined error originating from its two components, and $\varpi$ is the parallax from {\it Gaia} DR2. These cuts are designed to balance data quality and sample size. They do not introduce \emph{explicit} kinematic biases, which is necessary for our analysis below.
While the main sample used in our study uses white dwarfs within 250 pc, we will also occasionally use a subsample of white dwarfs located within 150 pc to clearly show number density enhancements.
\\
\\

\subsection{Kinematic and physical parameters of white dwarf}
\label{sec:quantity definitions}

Our analysis requires white dwarf absolute magnitude, color index, and the two components of transverse velocity $\bvT=(\vL,\vB)$ in Galactic longitude and latitude directions. Except for the color index $G_{\rm BP}$--$G_{\rm RP}$, which is directly read from the $\texttt{bp\_rp}$ column in {\it Gaia} DR2, we derive the other quantities in the following way:
\begin{align}
    M_G&=G+5\log(\varpi/\rm{mas}) -10\,,\\
    \vL &= \frac{\mu_{\rm L}-(A\cos 2l + B )\cos b}{\varpi}\,,\\
    \vB &= \frac{\mu_{\rm B}+A\sin 2l \sin b \cos b}{\varpi}\,,
\end{align}
where $G$ and $\varpi/\rm{mas}$ are read from {\it Gaia} DR2 columns  $\texttt{phot\_g\_mean\_mag}$ and $\texttt{parallax}$, $\mu_{\rm L}$ and $\mu_{\rm B}$ are converted from columns $\texttt{ra}$, $\texttt{dec}$, $\texttt{pmra}$, and $\texttt{pmdec}$ with the coordinate conversion function in the \textsc{astropy} package \citep{AstropyCollaboration_2013, AstropyCollaboration_2018}, and $A$ and $B$ are the Oort constants taken from \citet{Bovy_2017}. We do not correct for extinction because within the distance cut, extinction is in general tiny and there is no accurate estimate for it. To avoid the influence of hyper-velocity white dwarfs, we further impose a velocity cut:
\begin{align}
\label{eq:vT_cut}
    v_{\rm T}=\sqrt{\vL^2+\vB^2} < 250 {\rm\,km\,s^{-1}}\,.
\end{align}
We point out that {\it Gaia} does not provide any useful radial velocity information for white dwarfs as they have no spectral lines in the 845--872 nm wavelength range of {\it Gaia}'s spectrometer \citep{GaiaCollaboration_2016}.

We then derive white dwarf photometric isochrone ages and masses from the H--R diagram coordinates:
\begin{equation}
    (G_{\rm BP}\text{--}G_{\rm RP}, M_G)\rightarrow (\tau_\text{phot}, m_\text{WD})\,, 
\end{equation}
based on a single-star evolution scenario and white dwarf cooling models\footnote{We have made this tool a publicly available python 3 module on \WDmodelsurl}. We estimate the main-sequence (MS) ages with an initial--final mass relation \citep{Cummings_2018} and the relation between pre-WD time and main-sequence mass from \citet{Choi_2016} for non-rotating, solar-metallicity stars. For high-mass white dwarfs, the pre-WD ages are negligible. As for white dwarf cooling, we use a table of synthetic colors for pure-hydrogen atmosphere \citep{Holberg_2006, Kowalski_2006, Tremblay_2011} and a grid of cooling tracks for C/O-core white dwarfs with ``thick'' hydrogen layers \citep{Fontaine_2001}\footnote{\wdurl.}. In order to convert any H--R diagram coordinate into $\tau_\text{phot}$ and $m_{\rm{WD}}$, we linearly interpolate between grid points. Stellar models show that in the single-star-evolution scenario, white dwarfs with a mass higher than about 1.05--1.10~$M_\odot$ have oxygen+neon (O/Ne) cores \citep[e.g.,][]{Siess_2007,Lauffer_2018}. So, we combine the cooling tracks of $m_\text{WD}\leq1.05\,M_\odot$ C/O white dwarfs with the four cooling tracks of $m_\text{WD}\geq1.10\,M_\odot$ O/Ne white dwarfs \citep{Camisassa_2019}. 

The O/Ne white dwarf model only gives slightly lower mass estimates than the C/O white dwarf model (e.g., 1.08--1.23~$M_\odot$ in the combined model corresponds to 1.10--1.28~$M_\odot$ in the C/O-only model), and their estimates of the photometric ages $\tau_\text{phot}$ are similar for the white dwarfs we are interested in ($\tau_\text{phot}<$ 3.5 Gyr). Switching between thick-hydrogen, thin-hydrogen, and helium atmosphere \citep{Bergeron_2011} models does not significantly change the photometric-age estimate of our sample either. 

\subsection{Mass, age, and Q-branch selection}
\label{sec:sample selection}

In Figure~\ref{fig:WD_HR_70} we show the selected white dwarfs on the H--R diagram. In the top right panel, we use the 150 pc sample to show the density distribution with a higher contrast.
The Q branch is a factor-two enhancement at around $-0.4 <G_{\rm BP}\text{--}G_{\rm RP}< 0.2$ and $M_G=13$. In the main panel, we show our main sample within 250 pc, color-coded by their transverse velocities $v_{\rm T}$ with respect to the local standard of rest. We adopt $(U_\odot,V_\odot,W_\odot)=(11, 12, 7)$ km s$^{-1}$ \citep{Schonrich_2010} to correct for the solar reflex motion. We emphasize the fast white dwarfs ($v_{\rm T}>$ 70 km s$^{-1}$) in Figure~\ref{fig:WD_HR_70} with larger dots: they are very likely thick-disk stars. We also plot a grid of photometric age $\tau_\text{phot}$ and mass $m_\text{WD}$ derived from the combined O/Ne-core and C/O-core white dwarf cooling model. Cooling tracks are the curves with constant $m_\text{WD}$. White dwarfs with different birth times form a `white dwarf cooling flow' on the H--R diagram as they move along their cooling tracks. 

We focus on the mass range where the Q branch is most prominent. To maximize sample size and minimize the contamination from standard-mass helium-atmosphere white dwarfs (the B branch), we impose the following photometric age and mass cuts:
\begin{align}
    \text{0.1 Gyr} < &\,\tau_\text{phot} < \text{3.5 Gyr},\\
    \text{1.08 }M_\odot<&\,m_\text{WD}<\text{1.23 }M_\odot,
\end{align}
where $m_\text{WD}$ is derived from the combined cooling model for O/Ne and C/O white dwarfs. This mass range corresponds to 1.10--1.28~$M_\odot$ in the C/O-only cooling model. In total, 1070 white dwarfs are selected by these criteria\footnote{A catalog of all selected white dwarfs is available on VizieR and on the website: \schengurl}. In this region, the Q branch divides the white dwarf cooling flow into three segments: the early, Q-branch, and late segments, as shown in Figure~\ref{fig:WD_HR_70}. We define the Q-branch segment by
\begin{equation}
    |M_G-1.2\times\texttt{bp\_rp}-13.2|<0.2
\end{equation}
in addition to the previous photometric-age and mass cuts.

\section{An extra cooling delay on the Q branch}
\label{sec:extra cooling delay}

\subsection{Evidence from the photometric-age distribution}
\label{sec:enhancement}

\begin{figure}
    \centering
    \includegraphics[width=0.95\columnwidth]{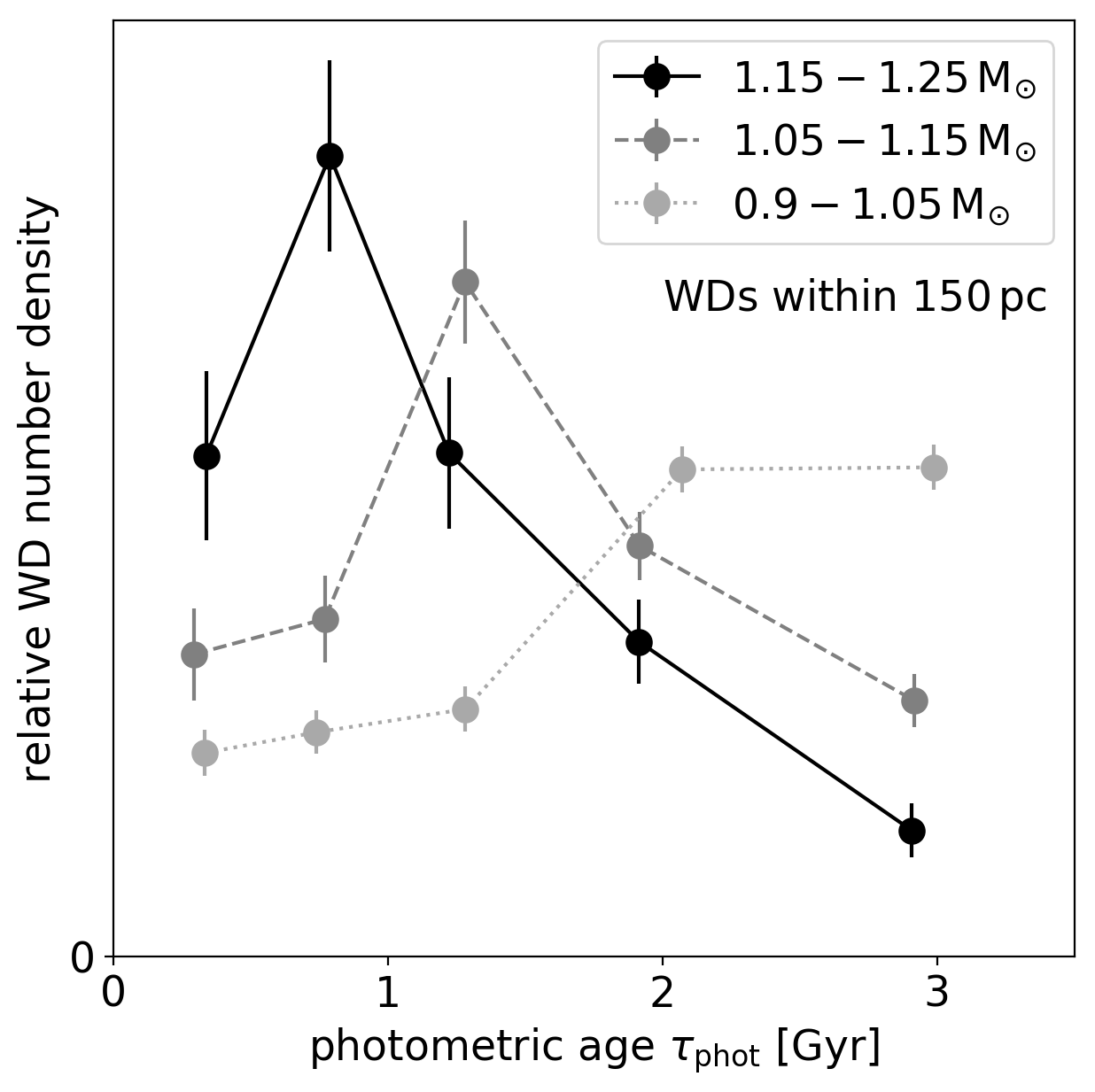}
    \caption{The normalized photometric age $\tau_\text{phot}$ distribution of high-mass WDs in three consecutive mass ranges. The mass-dependent peaks trace the position of the Q branch. Crystallization should not produce any peak on this plot, because the $\tau_\text{phot}$ is calculated from a model including crystallization effects; the completeness stays high for at least 1 Gyr after the peaks in each mass range, so these mass-dependent peaks cannot be explained by a peak in the star formation history or by incompleteness. Therefore, there must be an extra cooling delay piling up WDs on the Q branch.
    \label{fig:photometric age distribution}}
\end{figure}

\begin{figure*}[t]
    \centering
    \includegraphics[width=.458\textwidth]{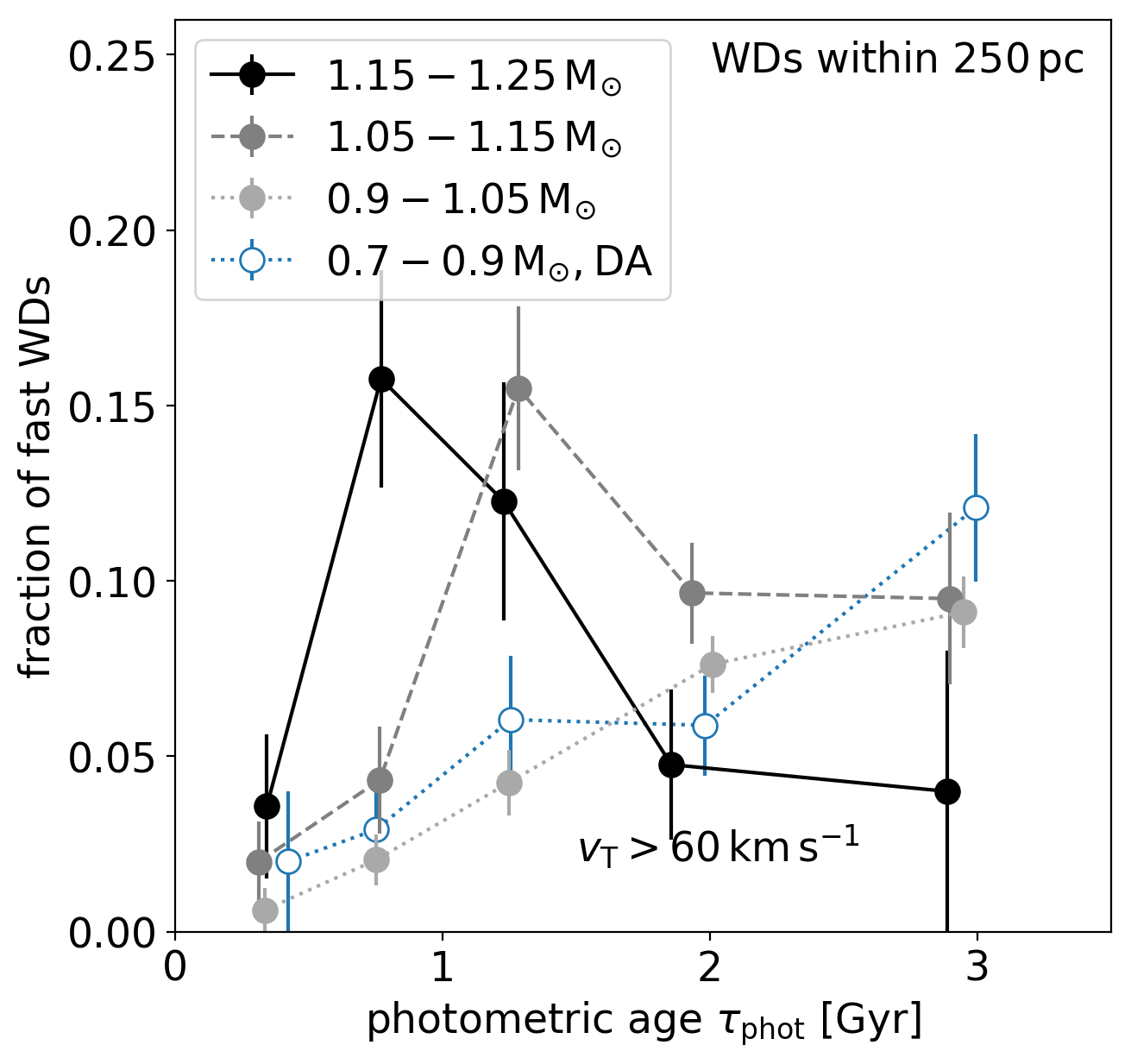}
    \includegraphics[width=.475\textwidth]{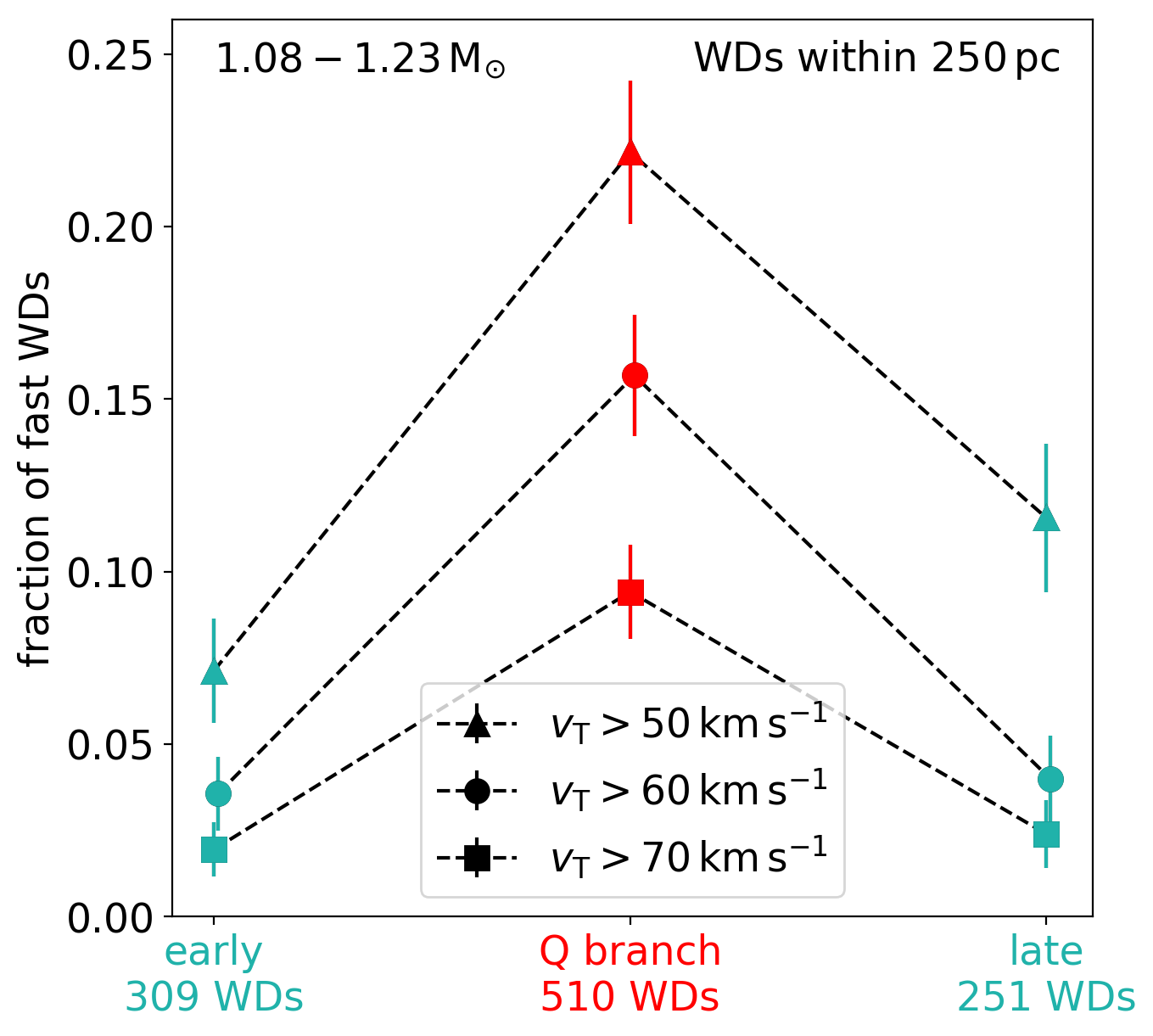}
    \caption{The fraction of fast WDs in different mass ranges (\emph{left panel}) and for different velocity cuts (\emph{right panel}). There are significantly more fast-moving WDs on the Q branch than both before and after it in terms of photometric age. According to the age--velocity-dispersion relation (AVR), fast WDs are old. As argued in Section~\ref{sec:velocity distribution}, this high fraction of fast WDs on the Q branch can only be explained by a subset of WDs experiencing an extra cooling delay on the Q branch.
    \label{fig:fast WD}} 
\end{figure*}

As argued by \citet{Tremblay_2019}, an enhancement not aligned with mass or age grid, such as the Q branch, should be produced by a slowing down (and therefore a delay) of white dwarf cooling. Such a cooling delay creates a `traffic jam' in the white dwarf flow, and the Q branch is a snapshot of this traffic jam. \emph{Is crystallization alone enough to explain the cooling delay on the Q branch?} If it is, then the distribution of photometric age $\tau_\text{phot}$ derived from a cooling model including crystallization effects should no longer carry signatures of the Q branch. However, observations lead to the antithesis. In Figure~\ref{fig:photometric age distribution} we show the distribution of $\tau_\text{phot}$ in three mass ranges: there is a mass-dependent enhancement tracing the Q branch, which is consistent with the observation by \citet{Tremblay_2019} that the pile-up is higher and narrower than what the standard cooling model predicts. Evolutionary delays from binary interactions or a peak in star formation rate cannot explain this mass-dependent enhancement either. Therefore, an extra cooling delay in addition to crystallization effects (latent heat and phase separation) must exist.

\subsection{Evidence from the velocity distribution}
\label{sec:velocity distribution} 

Observations show that the velocity dispersion of disk stars in the Milky Way is related to the stellar age $\tau$: older stars have higher velocity dispersion than younger stars \citep[e.g.,][]{Holmberg_2009}. So, the transverse velocity $\bvT$ derived from {\it Gaia} DR2 can be used as a `dynamical' indicator of the true stellar age $\tau$. For the Milky Way thin disk, the dispersion of transverse velocity approximately follows a power law increasing from about 25~km~s$^{-1}$ at 1.5~Gyr to 55~km~s$^{-1}$ at around 6--8~Gyr \citep[e.g.,][]{Holmberg_2009}; for the thick disk, the dispersion is about 65~km~s$^{-1}$ \citep[e.g.,][]{Sharma_2014}. Given this age--velocity-dispersion relation (AVR), we observe two anomalous things in the velocity distribution of the Q branch:
\begin{itemize}
    \item There is a strong excess of white dwarfs with $v_{\rm T}>$ 70~km~s$^{-1}$ in the Q-branch segment, as shown in Figure~\ref{fig:WD_HR_70}. According to the age--velocity-dispersion relation (AVR) mentioned above, these fast white dwarfs should be old stars. Given that the photometric age on the Q branch is only 0.5--2~Gyr, these white dwarfs must have experienced an extra cooling delay for several billion years. In the left panel of Figure~\ref{fig:fast WD} we show that the excess of fast white dwarfs in the Q-branch segment is clear for $m_\text{WD}>1.05\,M_\odot$; in the right panel we show that this excess is is observed for a variety of velocity cuts.
    \item The fraction of fast white dwarfs in the \emph{late} segment is \emph{lower} than that in the Q-branch segment. This is anomalous, because white dwarfs in the late segment should be older than those in the Q-branch segment, as long as all white dwarfs follow the same cooling law. The only way to create such a reverse of fraction is to have more than one white dwarf population with distinct cooling behaviors.
\end{itemize}

\subsection{A two-population scenario of the extra cooling delay}
\label{sec:two parameter description}

\begin{figure}
    \centering
    \includegraphics[width=\columnwidth]{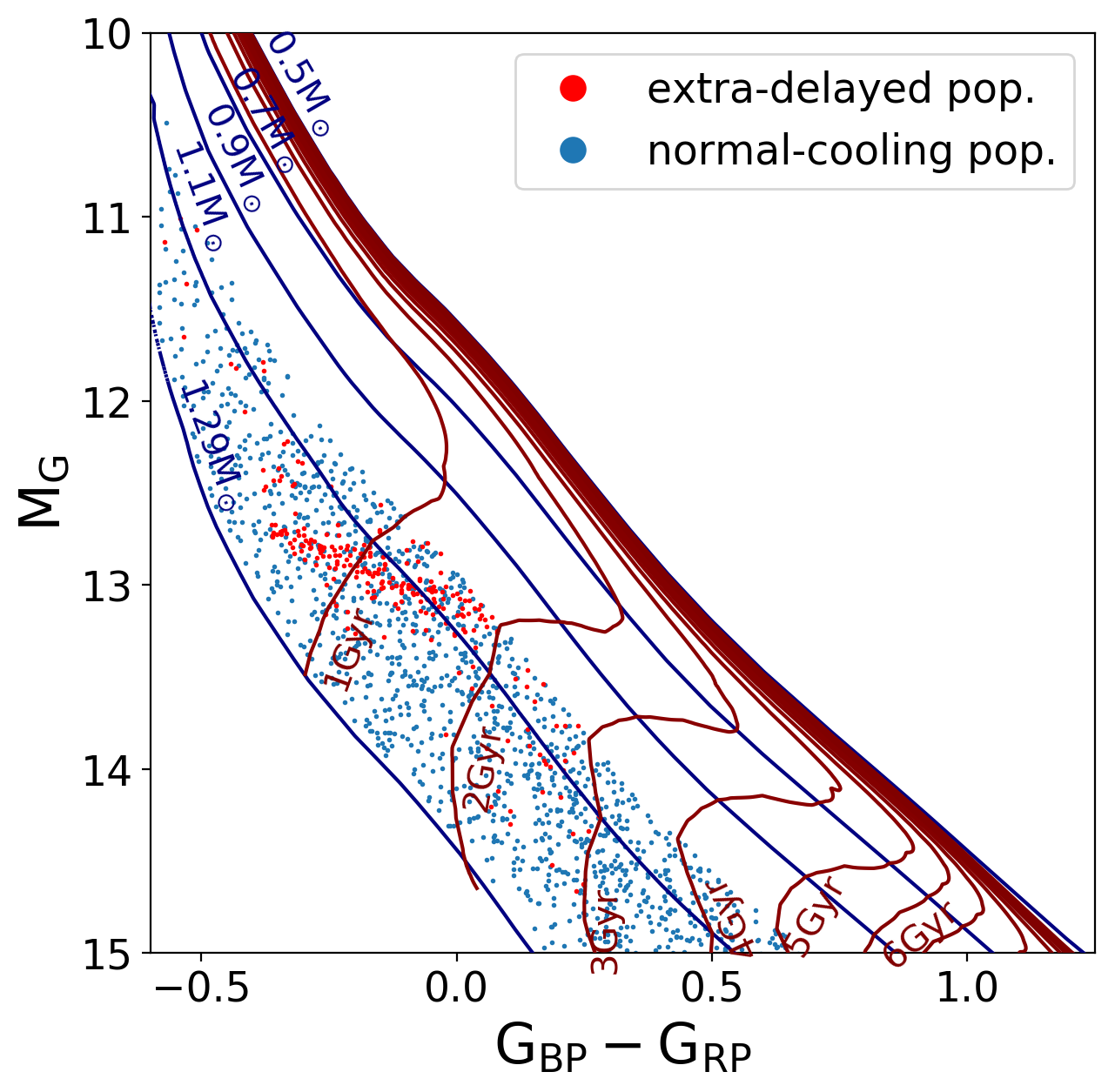}
    \caption{An illustration of the two-population scenario of the Q branch: a normal-cooling population (blue dots), and a population with the extra cooling delay (red dots). The number density of WDs on the H--R diagram is determined by the cooling rate, because WDs accumulate where the cooling rate is low. Here, we use the best-fitting values ($f_\text{extra}=$~6\% and $t_\text{extra}=$~8 Gyr) from our later analysis to generate this mock H--R diagram. An animated version of this figure is available, where blue dots move with the normal cooling rate, while red dots (the extra-delayed population) move slowly on the Q branch. Each second in the animation corresponds to 1 Gyr in physical time, and the duration of this animation is 11 seconds. More animations can be found on the website: \schengurl.}
    \label{fig:two populations}
\end{figure}

The simplest scenario that can explain both the number enhancement and velocity anomaly on the Q branch is to have an extra-delayed population of white dwarfs in addition to a normal-cooling population. This scenario requires only two free parameters:
\begin{itemize}
    \item The fraction $f_\text{extra}$ of the extra-delayed population, and
    \item The length $t_\text{extra}$ of the extra cooling delay on the branch.
\end{itemize}
In Figure~\ref{fig:two populations} we illustrate this scenario by showing the  normal-cooling and extra-delayed populations on the H--R diagram. Before the Q branch, the extra-delayed population has no difference from the normal-cooling population. On the branch, the extra-delayed population has a slower cooling rate, which causes two effects: (1) its members pile up there, creating a number-density enhancement, and (2) the photometric ages $\tau_\text{phot}$ of its members start to seem younger than their true ages $\tau$, creating an age discrepancy. After the branch, the number-density enhancement disappears, but the age discrepancy remains. A detailed parameterization of this scenario is presented in \ref{app:extra-cooling-delay scenario}. To create the observed reversal of fast white dwarf fraction in the Q-branch and late segment, the extra-delayed population must have a high number-density contrast between these two regions, which requires that the population fraction $f_\text{extra}$ be small and the delay time $t_\text{extra}$ long.

\section{Quantitative analysis}
\label{sec:quantitative inference}

Having shown qualitatively the existence of an extra cooling delay on the Q branch, we now attempt to quantify its properties. We build a statistical model that
(i) includes double-WD mergers, (ii) uses an anisotropic AVR, and (iii) makes use of the full constraining power of the observations.

\subsection{Merger products among high-mass white dwarfs}
\label{sec:double WD merger products}

Simulations of binary evolution show that double-WD merger products may account for a considerable fraction of high-mass white dwarfs \citep[e.g.,][]{Toonen_2017, Temmink_2019}. These merger products also have a discrepancy between their true ages and photometric ages due to binary evolution. Therefore, in order to use the velocity distribution to quantitatively constrain the extra cooling delay, the merger population must be modeled simultaneously. Constraining the merger fraction is also of great interest as its value is still a matter of debate \citep[e.g.,][]{Giammichele_2012, Wegg_2012, Rebassa-Mansergas_2015, Tremblay_2016, Maoz_2018}. Therefore, we include the double-WD merger products in in our model and set their fraction as a free parameter.
\\
\\

\begin{deluxetable*}{c|ccc}[t]
    \tablecaption{\label{tab:populations}Delay scenarios of the three populations}
    \tablehead{
        \colhead{Population}&\colhead{single-star evolution}&\colhead{extra-delayed}&\colhead{double-WD merger products with  normal-cooling}\\
        \colhead{(abbreviation)}&\colhead{(s)} & \colhead{(extra)} & \colhead{(m)}
    }
    \startdata
    merger delay & no & yes or no (setup 1 or 2)& yes\\
    extra cooling delay & no & yes & no\\
    \hline
    early & 0 & $\Delta t_{\rm merger}$ or 0 & $\Delta t_{\rm merger}$ \\
    Q branch & 0 & $(\Delta t_\text{extra}+\Delta t_{\rm merger})$ or $\Delta t_\text{extra}$  & $\Delta t_{\rm merger}$ \\
    late & 0 & $(t_\text{extra}+\Delta t_{\rm merger})$ or $t_\text{extra}$ & $\Delta t_{\rm merger}$
    \enddata
    \tablecomments{For each population, the delay types are shown in the upper part of the table, and the total delay time $\Delta t=\tau-\tau_\text{phot}$ for each segment is shown in the lower part. $\Delta t$, $\Delta t_{\rm merger}$, and $\Delta t_\text{extra}$ are not single numbers but random variables following their distributions. They are used to calculate the distributions of true ages $\tau$ from photometric ages $\tau_\text{phot}$.}
\end{deluxetable*}

\subsection{Description of the model}
\label{sec:description of the model}

In our model, we consider two evolutionary delays: the extra cooling delay, and the merger delay. Accordingly, we consider three populations of white dwarfs with different combinations of the two delays:
\begin{itemize}
    \item A generic population of singly evolved white dwarfs that follows normal cooling, denoted by `s';
    \item A double-WD merger population\footnote{We only consider the double-WD mergers because in our mass range, other merger products such as those from MS--RG, MS--MS, and MS--WD mergers usually only have $<$~0.2~Gyr delay and therefore are indistinguishable from the singly evolved white dwarfs in terms of kinematics.} with systematic age offsets due to the merger delay and with a normal cooling, denoted by `m';
    \item A population with the extra cooling delay, denoted by `extra'.
\end{itemize}
Their delay scenarios are listed in Table~\ref{tab:populations}.
For simplicity, we only explore the two extreme situations for the extra-delayed population, where
\begin{itemize}
    \item Setup 1: \emph{all} members of the extra-delayed population also have the merger delay;
    \item Setup 2: \emph{no} members of the extra-delayed population have the merger delay.
\end{itemize}
The distribution function $p(\mathbfit{y})$ of observables $\mathbfit{y}$ for all white dwarfs can be written as a weighted average of the distribution for each population:
\begin{equation}
\label{eq:components}
    p(\mathbfit{y}) = f_{\rm s}\,p_{\rm s}(\mathbfit{y}) + f_\text{m}\,p_{\rm m}(\mathbfit{y}) + f_\text{extra}\,p_{\rm extra}(\mathbfit{y})\,,
\end{equation}
where the weight $f$ denotes the fraction of each population, satisfying $ f_{\rm s} + f_\text{m} + f_\text{extra}= 1\,$.

Our goal is to use observations to constrain two independent population fractions and the delay time of the extra cooling delay:
\begin{equation}
    f_\text{m}, f_\text{extra}, \text{ and }t_\text{extra}\,,
\end{equation}
the last of which is encoded in the distribution $p_{\rm extra}(\mathbfit{y})$.
We have two sets of observables $\mathbfit{y}$: the transverse velocities $\mathbfit{v}_{\rm T}$, and the photometric ages $\tau_\text{phot}$. They are connected by the AVR $p(\mathbfit{v}|\tau)$ and the delay scenario of each population (listed in Table~\ref{tab:populations}). The delay
\begin{equation}
\label{eq:tau}
    \Delta t \equiv \tau - \tau_\text{phot}\,
\end{equation}
includes contributions from the extra-cooling and/or the merger delays. We build a Bayesian model based on Equation~\ref{eq:components} to constrain the aforementioned parameters.
Our model is similar to that of \citet{Wegg_2012}, but we include the extra-delayed population and use a much larger sample. In addition, to avoid the need for modeling selection effects, we derive our constraints from the velocity distribution conditioned on observables other than velocity:
\begin{equation}
\label{eq:conditional PDF}
    p(\mathbfit{v}|\text{other observables}) = p(\bvT|\tau_\text{phot},m_\text{WD},l,b)\,.
\end{equation}
The details of this statistical technique and the Bayesian framework of our model are shown in \ref{sec:bayesian framework}.

The free parameters in our model include the population fractions $f_\text{m}$ and $f_\text{extra}$, the extra delay time $t_\text{extra}$, parameters for the AVR, and solar motion. Although constraints on the AVR and solar motion already exist, treating them also as free parameters can avoid potential systematic errors, and the comparison of our best-fitting values with the existing values allows us to check the validity of our method.

Below, we list the main assumptions and simplifications in our model:
\begin{enumerate}
    \item We assume that upon entering the Q-branch segment, all members of the extra-delayed population suddenly slow down their cooling by a constant factor, and upon leaving the branch, the cooling rates suddenly resume, so that this extra cooling delay can be parameterized by just its length $t_\text{extra}$ and population fraction $f_\text{extra}$ (see \ref{app:extra-cooling-delay scenario}). The resulting delay-time distribution is described in Section~\ref{sec:delay distributions}.
    \item The velocity distribution of white dwarfs is a superposition of 3D Gaussian distributions as a function of age $\tau$, i.e. $p(\mathbfit{v}|\tau)= \mathcal{N}(\mathbf{v}_0(\tau), \mathbf{\Sigma}(\tau))$. The details of this Gaussian velocity model are shown in \ref{app:velocity model}.
    \item The true-age distribution of high-mass white dwarfs within 250~pc is uniform up to 11~Gyr, i.e. $\tau\sim U$[0, 11~Gyr].
    \item For the double-WD merger products, we follow \citet{Wegg_2012} and assume that the resulting white dwarf is reheated enough that its cooling age after the merger is almost equal to the photometric cooling age. We also assume a fixed delay-time distribution for double-WD mergers (see Section~\ref{sec:delay distributions}) and a parameterization of the AVR (see Section~\ref{sec:AVR}).
\end{enumerate}

\subsection{Delay-time distributions}
\label{sec:delay distributions}

The three white dwarf populations in our model are defined by their different delay signatures $\Delta t$, which concern the extra cooling delay $\Delta t_\text{extra}$ and double-WD merger delay $\Delta t_{\rm merger}$. The delay scenario of each population in each segment is listed in Table~\ref{tab:populations}.

The extra cooling delay $\Delta t_\text{extra}$ is built up on the Q branch. We adopt a uniform distribution $\Delta t_\text{extra}\sim U[0,t_\text{extra}]$ of this delay for white dwarfs in the Q-branch segment and a constant value in the late segment. Note that $\Delta t_\text{extra}$ is a random variable with a probability distribution, whereas $t_\text{extra}$, as a model parameter to be constrained, is the upper limit of $\Delta t_\text{extra}$.
In the Q-branch segment, we do not further distinguish if a white dwarf has just started or is about to complete their extra cooling delay, because the uncertainty of H--R diagram coordinate due to different atmosphere types and astrometric/photometric error is comparable to the width of the Q branch on the H--R diagram. In this case, a uniform distribution is a good and efficient approximation for $\Delta t_\text{extra}$.

The double-WD merger delay $\Delta t_{\rm merger}$ originates from the binary evolution before the merger. We refer to binary population synthesis results \citep[e.g.,][]{Toonen_2014} and approximate the delay by
\begin{equation}
    p(\Delta t_{\rm merger})\propto \Delta t_{\rm merger}^{-0.7}
\end{equation}
for $\Delta t_{\rm merger}>$~0.5 Gyr and zero for smaller $\Delta t_{\rm merger}$. Unlike the extra cooling delay, we do not set any free parameter for this merger-delay distribution.

\begin{figure}[t]
    \centering
    \includegraphics[width=\columnwidth]{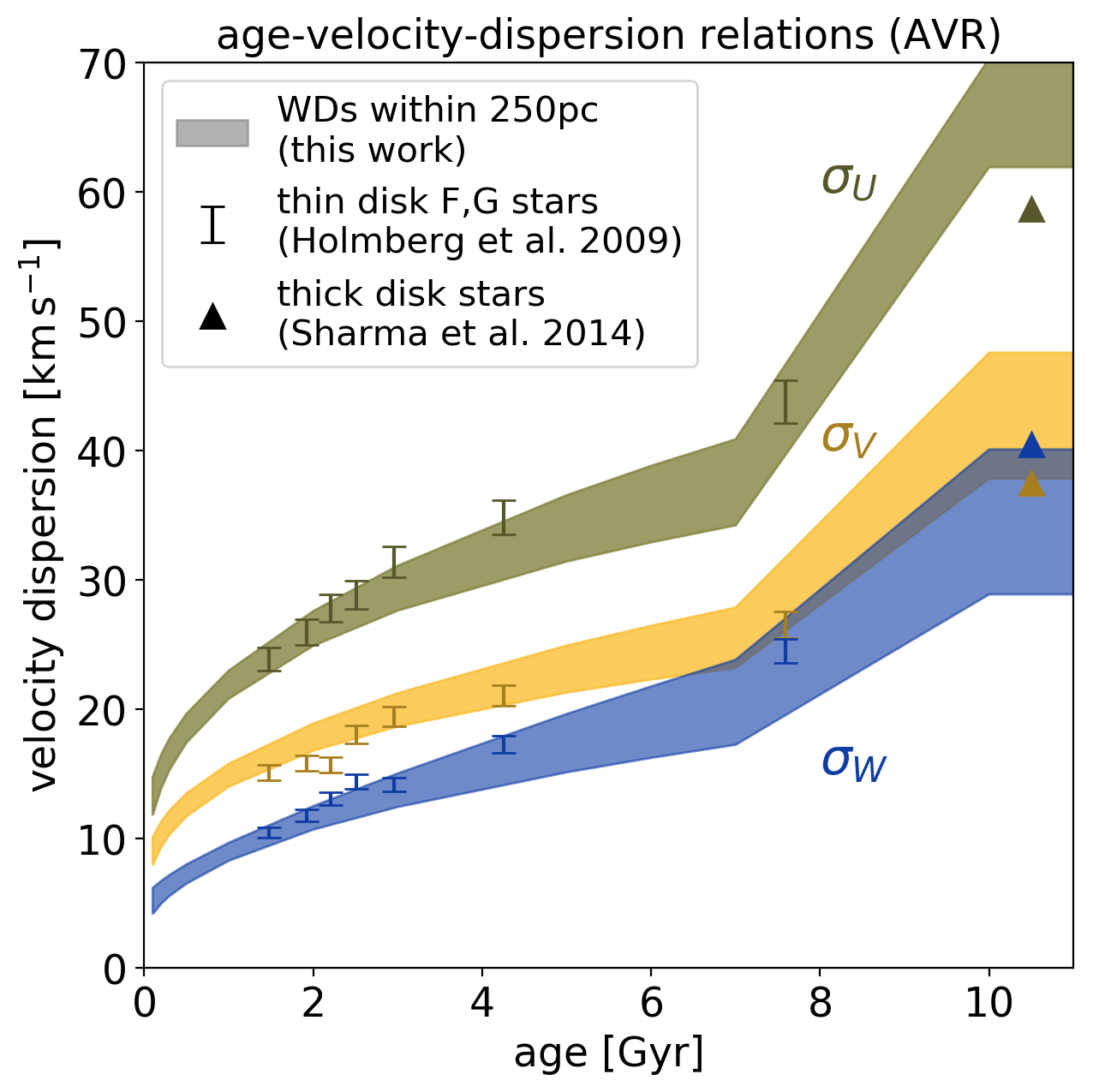}
    \caption{\label{fig:AVR} The comparison of AVRs constrained by this work and in the literature. The shaded regions show the 16th and 84th percentiles of the AVR posterior constrained by our high-mass WD sample. Symbols with error bars show the AVR measured for main-sequence stars by GCS and RAVE \citep{Holmberg_2009, Sharma_2014}. In our model, the $\tau<$~3.5 Gyr part of the AVR is mainly constrained by the normal-cooling WDs (population `s' and `m'), and the older part by the extra-delayed WDs. Note that the turnings at 7 and 10 Gyr reflect our parameterization of the AVR.}
\end{figure}

\begin{figure}[t]
    \centering
    \includegraphics[width=\columnwidth]{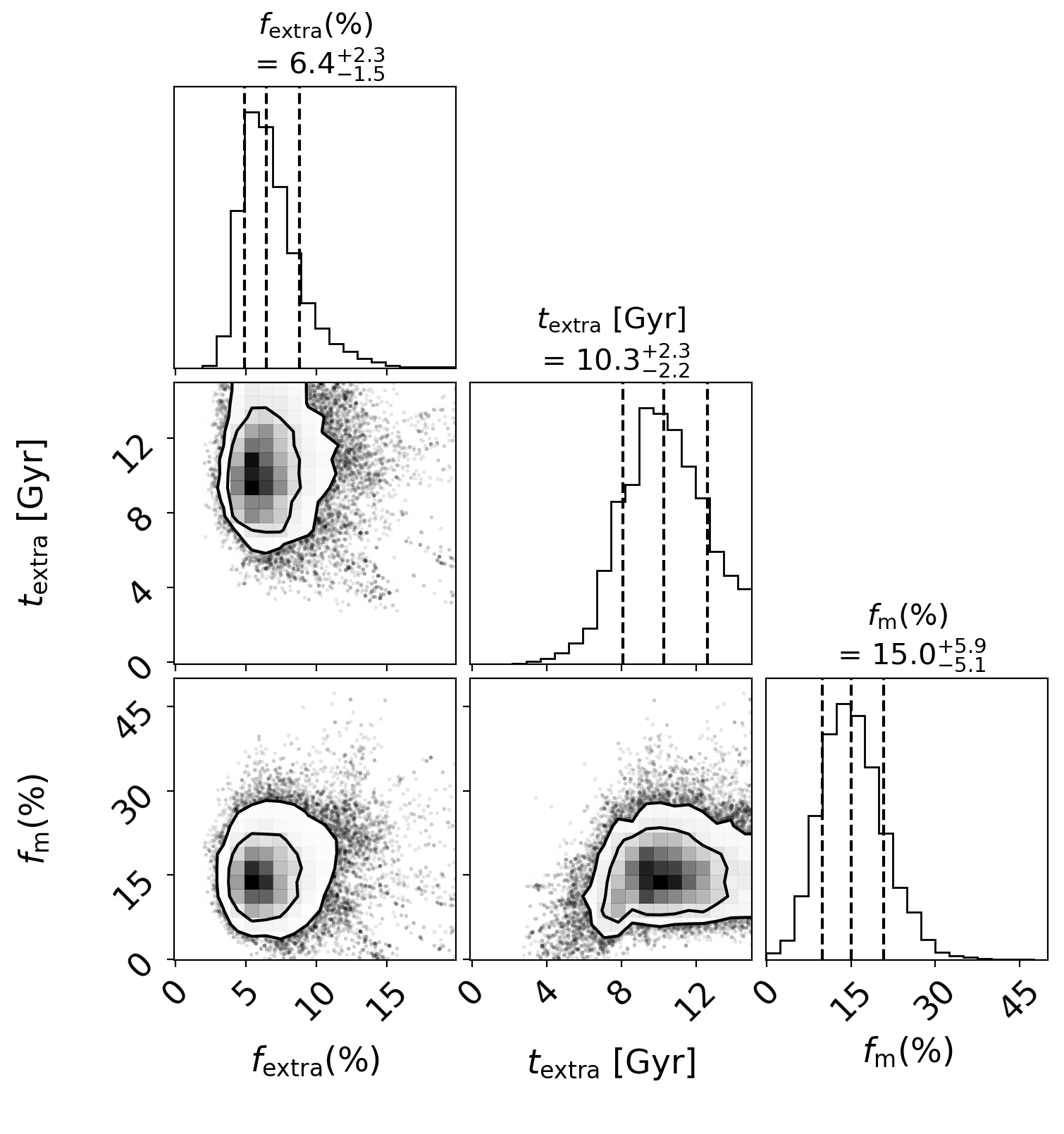}
    \caption{\label{fig:MCMC_main} The posterior distribution of the main parameters for setup 1. $f_\text{extra}$ is the fraction of extra-delayed population, $t_\text{extra}$ is the length of the extra cooling delay, and $f_\text{m}$ is the fraction of normal-cooling double-WD merger products. Note that in setup 1, it is $f_\text{extra}+f_\text{m}$ rather than $f_\text{m}$ that is the total fraction of merger products.}
\end{figure}

\subsection{Parameterization of the AVR}
\label{sec:AVR}

We define the $U$, $V$, $W$ axes as pointing toward $(l=0^{\circ},b=0^{\circ})$, $(l=90^{\circ},b=0^{\circ})$, and ($b=90^{\circ})$, respectively, and assume that the main axes of the Gaussian velocity distribution are aligned with these directions with dispersion $\sigma_{\rm U}(\tau)$, $\sigma_{\rm V}(\tau)$, and $\sigma_{\rm W}(\tau)$. Observations show that the AVR in each direction can be fit by a shifted power law. The power index of the in-disk components are around 0.35 and that of the $W$ component is around 0.5 \citep[e.g.,][]{Holmberg_2009, Sharma_2014}. For old stars including thick-disk members, the AVR is still a matter of debate \citep[e.g.,][]{Yu_2018, Mackereth_2019}. So, in each direction, we use a shifted power law to parameterize the AVR of the younger, thin-disk stars:
\begin{align}
\label{eq:AVR}
\sigma(\tau)=\sigma^{\tau=0}+\Delta\sigma^{0\rightarrow4}\times(\frac{\tau}{4})^\beta, &\;\;\;\;\tau<\text{7 Gyr}\,,
\end{align}
and we use a constant value $\sigma^{\rm thick}$ to represent the velocity dispersion of stars older than 10~Gyr (thick-disk stars); in between 7 and 10~Gyr, we linearly interpolate the values from the two ends to reflect the increasing fraction of thick-disk stars. The shape of the AVR with our parameterization is shown in Figure~\ref{fig:AVR}. The ratio of the two in-disk components $\sigma_{\rm V}$ and $\sigma_{\rm U}$ should be a constant for a local sample \citep[e.g.,][]{Binney_2008}, so we set $\sigma_{\rm V}(\tau) = k\,\sigma_{\rm U}(\tau)$. As the assumption for the velocity distribution to be Gaussian gradually breaks down when $\sigma_{\rm U}$ increases, we allow the ratio $k$ to be different for the thin and thick disks. Thus, we use in total 10 parameters to model the anisotropic AVR: two initial velocity dispersions $\sigma_{\rm U,W}^{\tau=0}$, two dispersion increases $\Delta\sigma_{\rm U,W}^{0\rightarrow4}$ between 0 and 4~Gyr, two power indices $\beta_{\rm U,W}$, two thick-disk dispersions $\sigma_{\rm U,W}^{\rm thick}$, and two in-disk dispersion ratios $k^{\rm thin}$ and $k^{\rm thick}$. The best-fitting values of these parameters can be checked against existing estimates presented in the literature.

\begin{figure*}[t]
    \centering
    \includegraphics[width=0.9\textwidth]{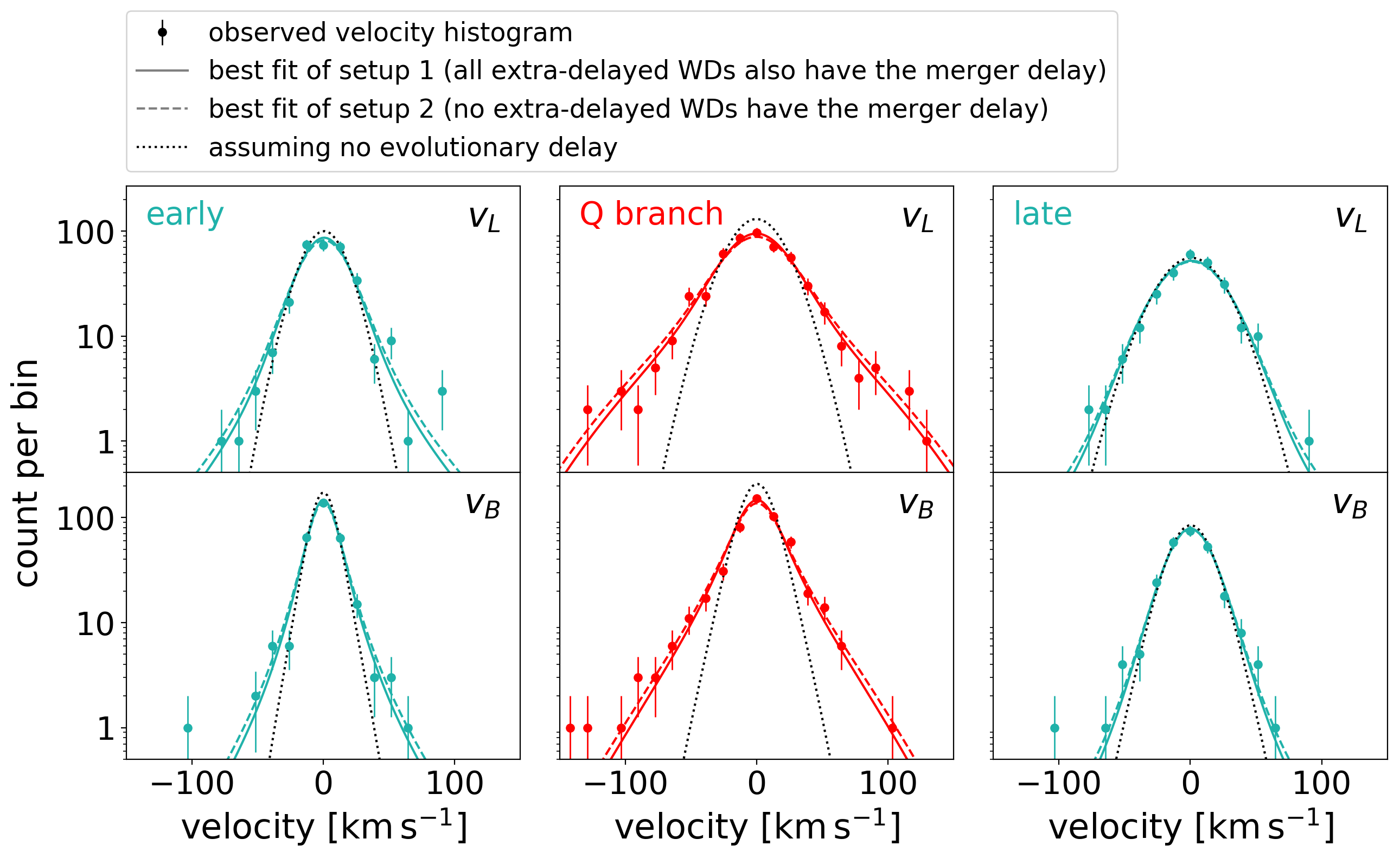}
    \caption{\label{fig:GOF} The observed and modeled velocity distributions. $\vL$ and $\vB$ are the Galactic longitude and latitude components of the transverse velocity. For the observed distribution, we present the histogram between -150 and 150 km s$^{-1}$ with 23 bins and the Poisson error of each bin. Note that the y-axis is in logarithmic scale. The solid and dashed curves, which are not very different, are the velocity distributions of the best-fitting models under setup 1 and 2, respectively. Both models fit the observations quite well. The dotted curves are the velocity distributions when no white dwarf has the extra cooling delay or merger delay. Its discrepancy to observed histograms shows the necessity of the two delays.}
\end{figure*}

\section{Results}
\label{sec:results}

To constrain the extra cooling delay properties and merger fraction, we feed our Bayesian model with the 1070 white dwarfs selected in Section~\ref{sec:data}. We use the Markov chain Monte Carlo (MCMC) sampler \texttt{emcee} \citep{Foreman-Mackey_2013} to obtain the posterior distribution of the parameters. Details of the settings are described in \ref{app:MCMC}.

\subsection{Constraints on the main parameters}
\label{sec:constraints}

In Figure~\ref{fig:MCMC_main} we present the constraints we obtain for the parameters of interest: $f_\text{extra}$, $t_\text{extra}$, and $f_\text{m}$. We find that the extra-delayed population fraction is
\begin{eqnarray}
\label{eq:f_extra}
    f_\text{extra}&=&6.4^{+2.3}_{-1.5}\,\%\text{~ (setup 1)}\nonumber\\
    &=&9.2^{+4.4}_{-2.7}\,\%\text{~ (setup 2),}
\end{eqnarray}
and the length of the extra cooling delay is
\begin{eqnarray}
\label{eq:t_extra}
    t_\text{extra}&\geq&\text{8 Gyr~ (setup 1)}\nonumber\\
    &\geq&\text{10 Gyr~ (setup 2).}
\end{eqnarray}
These constrains confirm our qualitative conclusion that $f_\text{extra}$ is small and $t_\text{extra}$ is long in Section~\ref{sec:two parameter description}. We point out that the difference of $t_\text{extra}$ in the two setups is exactly where the peak of the merger-delay distribution is located (2~Gyr), which is expected. The lower limit for $t_\text{extra}$ slightly depends on the parameterization of the AVR: if we adopt a younger thick-disk age \citep[7--11~Gyr instead of 9--11~Gyr,][]{Mackereth_2019}, this lower limit will also decrease by about 1--2~Gyr. $t_\text{extra}$ may be overestimated for the fact that at a high level of dispersion, the velocity distribution is not exactly Gaussian. But it is unlikely that we overestimate $t_\text{extra}$ too much, because we set the thick-disk dispersion as a free parameter. We have also checked that reasonable variations of the input delay-time distribution of the mergers \citep[e.g.,][]{Toonen_2012} do not change these two constraints significantly.

The fraction of merger products \emph{without} the extra cooling delay is found to be $f_\text{m}=15^{+6}_{-5}\,\%$ and $20^{+6}_{-5}\,\%$ (setups 1 and 2). Therefore, the total fraction of double-WD merger products is
\begin{align}
\label{eq:f_m}
    f_\text{extra}+f_\text{m}&=22^{+7}_{-5}\,\%\text{~ (setup 1)}\nonumber\\
    f_\text{m}&=20^{+6}_{-5}\,\%\text{~ (setup 2)}
\end{align}
among 1.08--1.23~$M_\odot$ white dwarfs. This total fraction is mainly constrained by the fast white dwarfs in the early segment (where the two setups do not differ from each other), so the constraints on this fraction under setups 1 and 2 are similar. A more detailed analysis of the merger products among high-mass white dwarfs is presented in \citet{Cheng_2019b}.

Finally, we calculate the contribution of the extra-delayed population in the Q-branch segment according to the above fractions:
\begin{align}
    F_\text{extra}&=\frac{f_\text{extra}(t_\text{extra}/\Delta t_\text{branch}+1)}{1+f_\text{extra}t_\text {extra}/\Delta t_\text{branch}}\nonumber\\
    &= (47\pm8)\,\%\;,
\end{align}
where $\Delta t_\text{branch}$ is the width of the Q branch (see \ref{app:extra-cooling-delay scenario}). This fraction, derived purely from the velocity information, is consistent with the factor $\sim$2 estimate obtained directly from the number-density enhancement in  Figure~\ref{fig:photometric age distribution}, which confirms that our extra-cooling-delay scenario is a good phenomenological model for the Q branch.

\subsection{Validation of the model}
\label{sec:validation}

To further validate our model and results, we first check our constraints on the nuisance parameters. For the solar motion we obtain
\begin{equation}
    (U,V,W)_\odot=(10.3\pm1.0, 7.3\pm1.0, 6.7\pm0.5)\text{~km s}^{-1},
\end{equation}
which is consistent with the measurement of \citet{Rowell_2019} based on mainly standard-mass white dwarfs. Our values of $U_\odot$ and $W_\odot$ are also consistent with the results of \citet{Schonrich_2010}. The discrepancy of the $V_\odot$ measurement comes from the different treatments of the asymmetric drift and is beyond the scope of this paper.

Figure~\ref{fig:AVR} shows our constraint on the white-dwarf AVR, which is consistent with the AVR of thin- and thick-disk main-sequence stars \citep{Holmberg_2009, Sharma_2014}.
Removing either the extra-delayed population or the merger population leads to unreasonably higher AVRs. 
Before {\it Gaia} DR2 came out, \citet{Anguiano_2017} reported an unexpectedly high AVR for young white dwarfs (their figures 21 and 22), without considering the extra cooling delay or merger delay in their age estimate. This unreasonably high AVR is exactly what we see when we remove the extra-delayed and/or merger population from our model.
Thus, we verify that both the extra cooling delay and the merger delay are necessary.

To check the goodness of fit, we compare the observed and modeled velocity distributions in Figure~\ref{fig:GOF}. Our best-fitting models (in both setups 1 and 2) provide good characterizations of the observed velocity distribution in all the early, Q-branch, and late segments. Adopting a different star formation history introduces no significant changes to our results. We test both a linearly decreasing star formation rate with a five-time higher star formation rate in the past, and a star formation history with a bump at 2.5 Gyr ago \citep[e.g.,][]{Mor_2019}, and find that the changes in best-fitting values are smaller than their uncertainties. This insensitivity to the assumed star formation history is expected because our model mainly uses the velocity information.

To further argue for our extra-delayed scenario against other explanations of the velocity anomaly, such as an accretion event of the Milky Way, we run a simple test where the velocities of fast white dwarfs on the Q branch are parameterized by only one Gaussian distribution. We find that the mean of the $U$ and $W$ components are consistent with zero, and the mean of $V$ is $-50\pm6$ km s$^{-1}$. Moreover, the $U$ component has a dispersion of $60\pm6$ km s$^{-1}$ and the ratio between the $U$ and $V$ dispersion is 0.60 $\pm$ 0.08. All of these values satisfy the relations for a disk in equilibrium \citep[e.g.,][]{Binney_2008}: asymmetric drift $\bar{V}=-\sigma_{\rm U}^2/(80\text{ km s}^{-1})$ and dispersion ratio $\sigma_{\rm V}/\sigma_{\rm U}=$ 0.67. It is unlikely for accreted stars to exactly reproduce the disk kinematics.

\section[The physics behind the extra cooling delay: 22Ne settling?]{The physics behind the extra cooling delay: \\$^{22}$Ne settling?}
\label{sec:22Ne}

In previous sections, we showed that a previously unreported cooling delay is required to explain the velocity distribution of white dwarfs on the Q branch.
Physically, this extra cooling delay requires an energy source satisfying the following conditions:
\begin{enumerate}
    \item It has a highly peaked effect on the Q branch;
    \item It is selective and applies to only $f_\text{extra}\sim$ 6\% of high-mass white dwarfs;
    \item It is powerful enough to create a $t_\text{extra}\sim$ 8~Gyr delay (in addition to crystallization delay and merger delay).
\end{enumerate}

These requirements are very demanding. For example, a higher energy release from latent heat or phase separation is ruled out because their effects are not peaked enough and they are not selective. Besides crystallization, another possible energy source in a white dwarf is the settling of $^{22}$Ne \citep{Isern_1991, Bildsten_2001}. Below, we show that $^{22}$Ne settling could account for the extra cooling delay.

\begin{figure}
    \centering
    \includegraphics[width=\columnwidth]{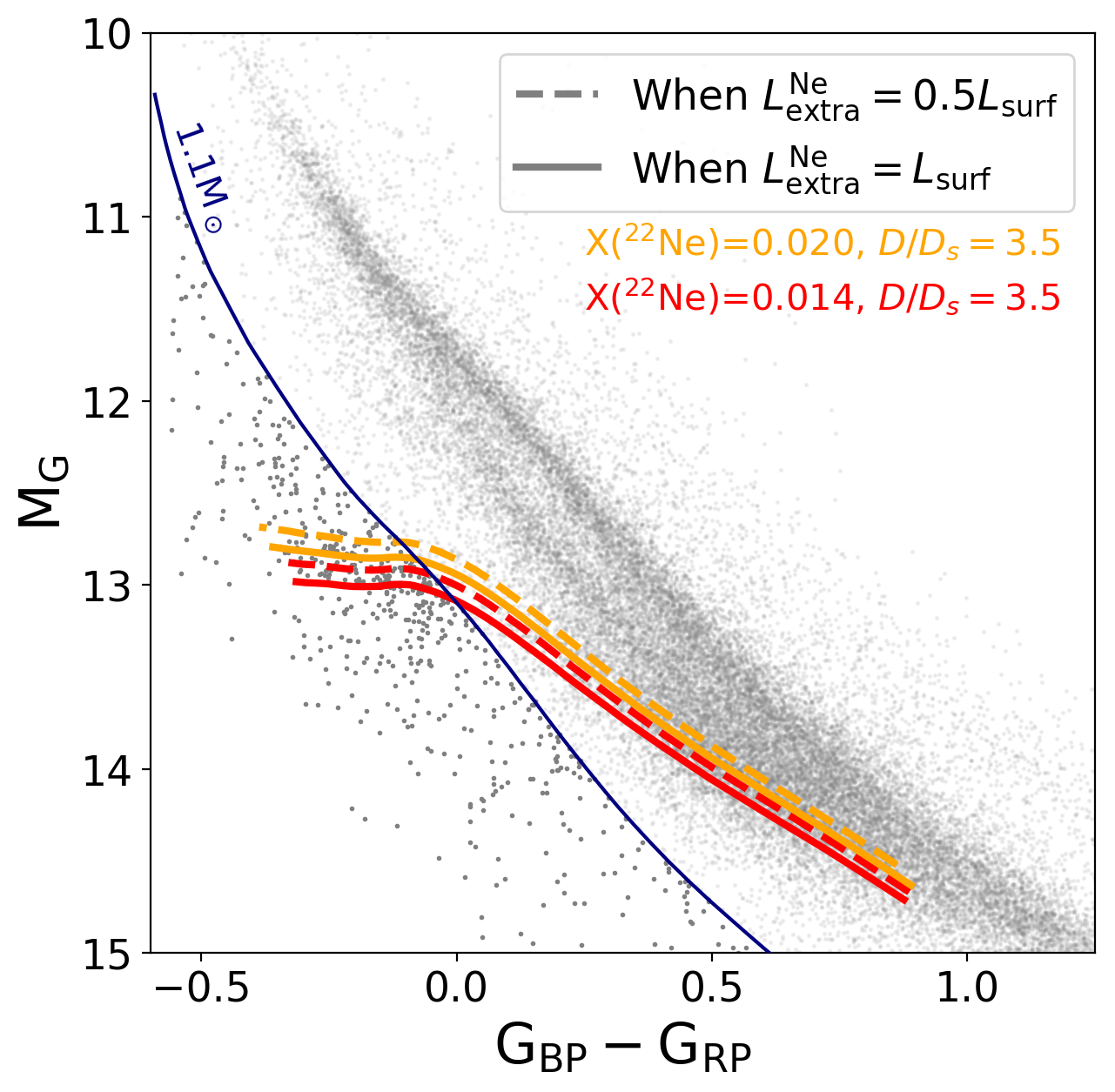}
    \caption{\label{fig:22Ne}The effective zone of $^{22}$Ne settling for C/O-core DA WDs with $D/D_s=$ 3.5, assuming no suppression from crystallization. $^{22}$Ne settling significantly delays the WD cooling between the solid and dashed curves, when $L_\text{extra}^\text{Ne}$ is close to $L_\text{surf}$. The colors represent two $^{22}$Ne abundances of WDs, corresponding to $\rm[M/H]=$ 0 and 0.15 in their progenitor stars. 
    We observe that the position, trend, and narrowness of the effective zone of $^{22}$Ne settling match the Q branch quite well.}
\end{figure}

Different from the large amount of $^{20}$Ne in O/Ne-core white dwarfs, the neutron-rich $^{22}$Ne is produced from C, N, and O in the core of the progenitor stars. At the hydrogen burning stage, the CNO cycle builds up the slowest reactant $^{14}$N, and at the helium burning stage, all $^{14}$N is converted into $^{22}$Ne. This leads to an abundance $X^\text{WD}_{^{22}\text{Ne}}\approx X^\text{star}_\text{CNO}\approx$ 0.014 for solar metallicity stars. Due to the additional two neutrons, $^{22}\rm{Ne}$ nuclei feel more downward force from gravity than the upward force from the electron-pressure gradient. So, they gradually settle down to the white dwarf center and release gravitational energy \citep{Bildsten_2001}.

Now, let us check if $^{22}$Ne settling satisfies the three requirements. We first emphasize that the delay \emph{effect} only depends on the \emph{fractional} contribution of the extra energy source to the white dwarf luminosity ($L_\text{extra}/L_\text{surf}$, see \ref{app:22Ne} for details). Therefore, to create a peaked effect, $L_\text{extra}$ need not be also peaked.

The luminosity of $^{22}$Ne settling ($L^\text{Ne}_\text{extra}$) relies on the $^{22}$Ne abundance, mass, and core composition of the white dwarf, and the inter-to-self-diffusion factor $D/D_s$, which is of order unity but not well-determined. As a white dwarf cools down, $L^\text{Ne}_\text{extra}$ does not change much, whereas $L_\text{surf}$ drops quickly with temperature \citep[e.g., Figure 2 of][]{Bildsten_2001}. So, if no suppression of $^{22}$Ne settling, the two luminosities will meet at some temperature. Around this meeting point is the effective zone of $^{22}$Ne settling, where the white dwarf cooling rate is influenced significantly. On the other hand, the meeting temperature is a function of white dwarf mass. We derive this temperature--mass relation in \ref{app:22Ne} and translate it into H--R diagram coordinates. In Figure~\ref{fig:22Ne} we show the results for $X(^{22}\text{Ne})=$ 0.014 and 0.020 ($\rm [M/H]=$ 0 and 0.15 in the progenitor stars), $D/D_s=3.5$, C/O-core white dwarfs. The effective zone of $^{22}$Ne settling is indeed highly peaked, and it matches the position and shape of the Q branch well.

Crystallization is a mechanism that may suppress the $^{22}$Ne settling by reducing its mobility in the plasma \citep[e.g.,][]{Bildsten_2001, Deloye_2002}. Therefore, in order to see a strong effect of $^{22}$Ne settling, $L^\text{Ne}_\text{extra}$ must be high enough to let the meeting point precede crystallization (see \ref{app:22Ne}). Because $^{22}$Ne settling favors C/O-core and previously metal-rich white dwarfs versus O/Ne-core and/or previously metal-poor white dwarfs, and crystallization sets in earlier in O/Ne-core white dwarfs than C/O-core white dwarfs, the delay effect of $^{22}$Ne settling is indeed selective. It is worth noting that high-mass C/O-core white dwarfs are believed not to be singly evolved \citep[e.g.,][]{Siess_2007, Lauffer_2018}, which means that if the extra cooling delay is really caused by $^{22}$Ne settling, then the extra-delayed white dwarfs should originate from double-WD mergers, i.e., our setup 1 is correct.

The gravitational energy of $^{22}$Ne stored in 1.0 and 1.2~$M_\odot$ white dwarfs ($Z=$ 0.02) are $6.8\times10^{47}$ and $1.5\times10^{48}$~ergs \citep{Bildsten_2001}; the surface luminosity of white dwarfs on the Q branch is $10^{-3.2}$ and $10^{-2.7}\,L_\odot$ for the two masses. If crystallization sets in later than this luminosity, $^{22}$Ne settling can stop their cooling for around 8.9 and 6.2~Gyr, respectively, close to our observational constraint for the extra cooling delay. Existing numerical simulations \citep{Deloye_2002, Garcia-Berro_2008, Althaus_2010, Camisassa_2016} give shorter delays (0.2--4.1~Gyr) for white dwarfs with even the highest possible $L^\text{Ne}_\text{extra}$. However, the delay time is sensitive to the choice of $D/D_s$ and temperature of crystallization, but existing models have only sparsely sampled the parameter space. Moreover, for the two-component C/O plasma, the updated phase diagram \citep{Horowitz_2010, Hughto_2012} suggests a much lower melting temperature than the widely used phase diagram of \citet[][]{Segretain_1993} and the naive prescription of using the same condition as in one-component plasma ($\Gamma=$ 178). This low melting temperature means a later crystallization, which can lengthen the delay of $^{22}$Ne settling.

In summary, we propose $^{22}$Ne settling as a promising candidate for the physical origin of the extra cooling delay. $^{22}$Ne settling has a more significant effect in C/O-core white dwarfs, which suggests that the extra-delayed white dwarfs are also merger products. To test our idea, detailed cooling models of high-mass C/O white dwarfs are needed.

\section{Discussion}
\label{sec:discussion}

In this section, we discuss two other observational features of the Q branch: the concentration of DQ white dwarfs, and the lack of wide-binary systems. Both of them support the idea that the extra-delayed white dwarfs may also be double-WD merger products, which has been suggested from the $^{22}$Ne-settling explanation.

\subsection{Concentration of DQ white dwarfs on the Q branch}
\label{sec:DQ}

\begin{figure}[t]
    \centering
    \includegraphics[width=\columnwidth]{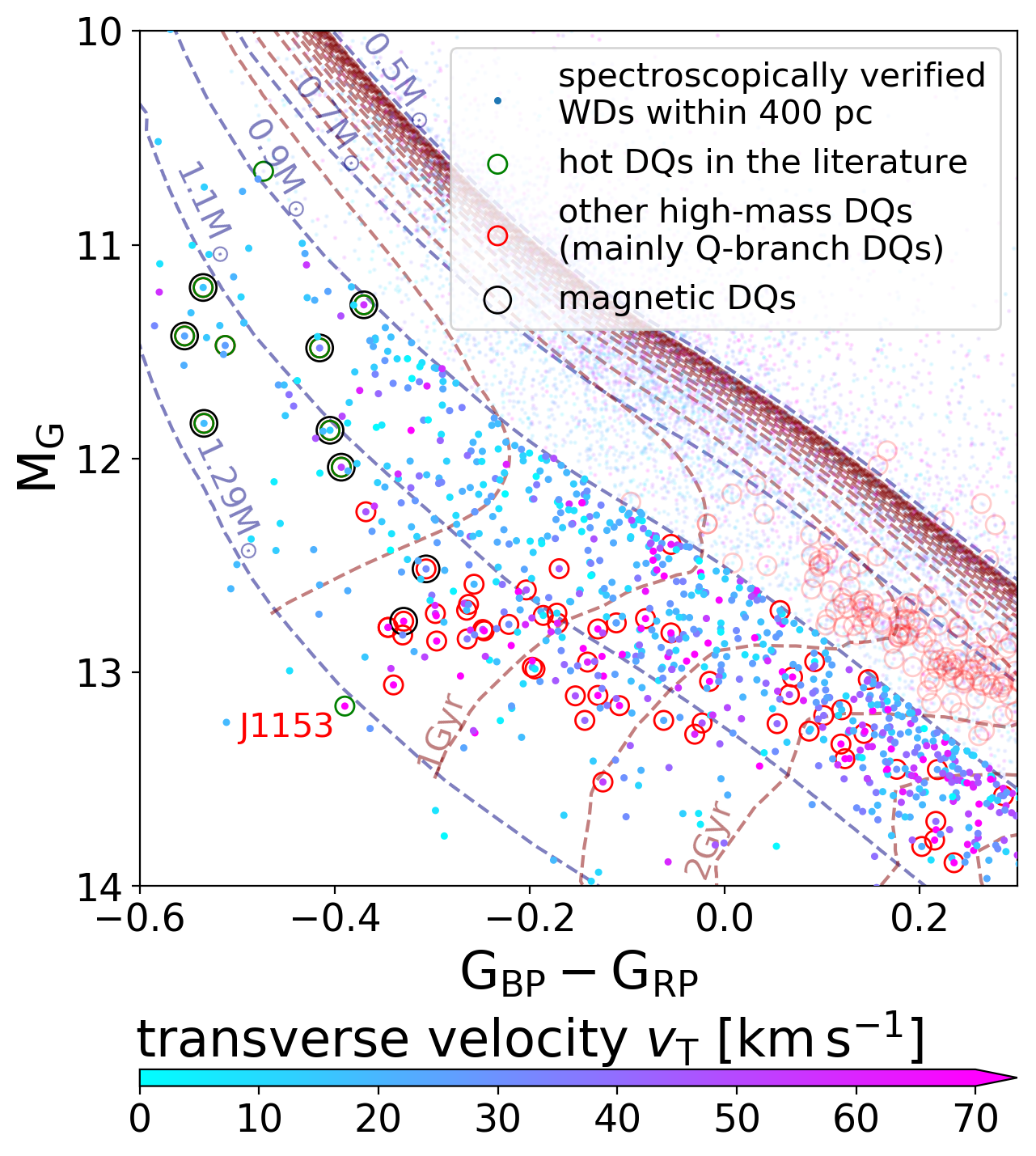}
    \caption{\label{fig:DQ}A part of the H--R diagram showing the spectroscopically verified WDs, with Q-branch DQ and hot-DQ white dwarfs marked by red and green open circles. The dots without circles are mostly DA white dwarfs. We estimate that half of the extra-delayed white dwarfs are DAs, because half of the fast-moving white dwarfs on the Q branch are DAs. We also mark the known magnetic DQs with larger black circles. Note that the mass range here ($>0.9\,M_\odot$) is larger than for the sample in our main analysis (1.08--1.23~$M_\odot$).}
\end{figure}

The Q branch is named after the presence of DQ-type white dwarfs \citep{GaiaCollaboration_2018a}. To explore this dimension, we cross-match our white dwarf sample with the Montreal white dwarf database, MWDD \citep{Dufour_2017}\footnote{\MWDDurl}.
We note that most high-mass DQs are concentrated on the branch (Figure~\ref{fig:DQ}) and the fraction of fast DQs on the branch is very high (Table~\ref{tab:DQ}). Therefore, all of these Q-branch DQs are likely to belong to the extra-delayed population. However, \emph{not all extra-delayed white dwarfs are DQs}. We estimate the fraction of DQs in the extra-delayed population to be
\begin{equation}
\label{eq:DQ fraction}
F_{\rm DQ}=19/(76F_\text{extra})=(53\pm16)\,\%\,,
\end{equation}
based on the total number of DQs and DAs in Table~\ref{tab:DQ}. Changing the distance limit of the sample does not influence this result much. It remains unclear but is of further interest to investigate the reason why half of the extra-delayed white dwarfs are DQs while the other half are DAs.

The DQs on the Q-branch are anomalous because the convection zone in a normal white dwarf with similar temperature is not deep enough to dredge up carbon \citep{Dufour_2005}. In a similar way, the hot-DQ white dwarfs discovered by \citet{Dufour_2007} are also abnormal. In Figure~\ref{fig:DQ} we show the distributions of the Q-branch DQs and hot-DQs on the H--R diagram. Below, we argue that although these two groups of DQs are observationally different, they may be related through an evolutionary relation.

The hot-DQs and Q-branch DQs appear to be different in some aspects. Hot-DQ white dwarfs are characterized by the high temperature ($>$18,000 K), highly carbon-dominant atmosphere \citep{Williams_2013}, high rate of having a magnetic field \citep{Dufour_2010, Dufour_2013}, high rate of being variable \citep[e.g.,][]{Dufour_2009, Dunlap_2010, Dufour_2011, Williams_2016}, and rarity \citep[e.g.,][]{Dufour_2008}. In contrast, the Q-branch DQ white dwarfs are concentrated on the Q branch, have helium-dominant atmospheres with 10$^{-4}$--10$^{-1}$ carbon \citep{Kepler_2015, Kepler_2016, Coutu_2019}, and have undetectable or no magnetic field (see Figure~\ref{fig:DQ}; a caveat for the magnetic field is that most hot-DQs have been examined with high-resolution spectroscopy, so their magnetic fields are more likely to be found). As for kinematics, hot-DQs are mildly faster than normal white dwarfs, which is an indication of being merger products \citep{Dunlap_2015}, whereas Q-branch DQs are much faster, which needs the long extra cooling delay to explain. \citet{Dunlap_2015} discussed one strange hot-DQ (SDSS J115305.47+005645.8 or J1153) with a very high proper motion. We note that J1153 has not been reported to have magnetic field or variability and lies on the Q branch (Figure~\ref{fig:DQ}), which means that J1153 can be classified as a Q-branch DQ. 

\begin{deluxetable}{c|cccccc}
\tablecaption{The statistics of velocity and spectral type of white dwarfs on the Q branch\label{tab:DQ}}
\tablehead{
    \colhead{250 pc spectro-} &\colhead{all} &\colhead{DQ} &\colhead{DA}\\
    \colhead{scopic sample} &\colhead{} &\colhead{} &\colhead{}
}
\startdata{
    all $v_{\rm T}$ & 76 & 19 & 53 \\
    \hline
    $v_{\rm T}>50{\rm\,km\,s^{-1}}$ & 23 & 8 & 14\\
    & $30\pm6\%$ & $42\pm15\%$ & $26\pm7\%$ \\
    \hline
    $v_{\rm T}>60{\rm\,km\,s^{-1}}$ & 16 & 7 & 8\\
   & $21\pm5\%$ & $37\pm14\%$ & $15\pm6\%$\\
    \hline
    $v_{\rm T}>70{\rm\,km\,s^{-1}}$ & 9 & 2 & 6\\
   & $12\pm4\%$ & $11\pm7\%$ & $11\pm5\%$\\
}\enddata
\tablecomments{The fraction of fast DQs is consistent with it belonging purely to the extra-delayed population.}
\end{deluxetable}

We now turn to the similarities between Q-branch DQs and hot-DQs: both of them have high masses, and both of them might have a merger origin. These two similarities raise a serious question: \emph{are the Q-branch DQs evolved from the hot-DQs?} We use number counts to explore this possibility. Hot-DQs are rare; based on a spectroscopically verified white dwarfs sample (as shown in Figure~\ref{fig:DQ}), we find that the fraction of hot-DQs is $8 / 203 =4.0\pm1.4\,\%$ in the region earlier than the Q branch ($\tau_\text{phot}<$ 0.5 Gyr, $m_\text{WD}>0.9\,M_\odot$). As a comparison, our estimate of the extra-delayed population fraction ($f_\text{extra}$) in this region is 6.4\% (Equation~\ref{eq:f_extra}) and about half of them are DQs (Equation~\ref{eq:DQ fraction}). So, these number counts are consistent with the scenario that hot-DQs are the evolutionary counterparts of Q-branch DQs.

In summary, based on the velocity distribution, we argue that all of Q-branch DQs belong to the extra-delayed population, and they account for $53\pm16\,\%$ of this population. In terms of observational properties, Q-branch DQs form a new class of DQ white dwarfs in addition to hot-DQs and the well-understood standard-mass DQs. However, number counts show that hot-DQs may evolve into Q-branch DQs, and both of them are likely to originate from double-WD mergers.

\subsection{Lack of wide binaries on the Q branch}
\label{sec:wide binary}

One additional way to test if the extra-delayed white dwarfs are also merger products is to check the wide binary fraction. The kick velocity of a few km s$^{-1}$ \citep[estimated from the results of][]{Dan_2014} from a merger may destroy many wide-separation binaries, making the wide-binary fraction lower. Because the extra-delayed population is significantly enhanced on the Q branch, if the extra-delayed white dwarfs are double-WD merger products, one would expect to see a lower wide-binary fraction on the Q branch.

We cross-match the wide binaries in {\it Gaia} DR2 \citep{El-Badry_2018, El-Badry_2019} with our high-mass white dwarf sample. In the early, Q-branch, and late segments, we find 5, 4, and 7 white dwarfs with wide-separation companions out of 309, 510, and 251 white dwarfs, respectively. So, the wide-binary fraction \emph{on} the Q branch is $0.8\pm0.4\,\%$, $2\sigma$ lower than the value \emph{off} the branch ($2.2\pm0.5\,\%$). If we assume that the extra-delayed population contributes no wide-binary system, then the wide-binary fraction of the normal-cooling populations \emph{on} the Q branch becomes $4/[510\times(1-F_\text{extra})]=(1.7\pm0.8)\,\%$, consistent with the \emph{off}-branch value $2.2\pm0.5\,\%$ within $1\sigma$. Therefore, the fraction of wide binaries provides additional support for the idea that the extra-delayed white dwarfs are double-WD mergers products.

\section{Conclusion}
\label{sec:conclusion}

The white dwarf H--R diagram derived from {\it Gaia} data has revealed a number-density enhancement of high-mass white dwarfs, called the Q branch. This branch coincides with the crystallization branch, but it is more peaked than what crystallization can create. Adding transverse-velocity information to the H--R diagram, we find a clear excess of fast white dwarfs on the Q branch (Figure~\ref{fig:WD_HR_70}). According to the age--velocity-dispersion relation (AVR) of Milky Way disk stars, these fast white dwarfs are much older than their photometric isochrone ages. Therefore, both the number count and velocity distribution suggest an extra cooling delay on the Q branch.

Motivated by these simple observations, we build a Bayesian model to quantitatively investigate this extra cooling delay. Because double-WD merger products also contribute to high-mass white dwarfs, we consider in our model both the extra cooling delay and the double-WD merger delay. Our model includes three white dwarf populations: one with no evolutionary delay, one with only the merger delay, and one with the extra cooling delay. We explore both situations in which all (setup 1) and none (setup 2) of the extra-delayed white dwarfs also have the merger delay. Our statistical model uses the discrepancy between the dynamical age inferred from transverse velocity and the photometric age to constraint the fraction of each white dwarf population and the length of the extra cooling delay. To eliminate selection effects, we model the conditional probability distribution of the transverse velocity of each white dwarf given its H--R diagram coordinate and spatial position. To avoid systematic errors of the model, we fit the solar motion and the anisotropic AVR together with the main parameters of interest.

We feed the model with 1070 high-mass white dwarfs (1.08--1.23~$M_\odot$, 0.1~Gyr $<\tau_\text{phot}<$ 3.5~Gyr, and $d<$ 250~pc) selected from {\it Gaia} DR2. Having checked that the AVR and solar motion parameters are all in agreement with standard values from the literature and that our best-fitting model provides a good fit to the observed velocity distribution, we find
\begin{enumerate}
    \item about 6\% of the high-mass white dwarfs experience an extra cooling delay that significantly slows down their cooling and makes them stay on the Q branch for about 8~Gyr;
    \item in the Q-branch region, an enhanced fraction (about a half) of the white dwarfs are extra-delayed due to the pile-up effect;
    \item half of the extra-delayed white dwarfs are DQs;
    \item as a byproduct of our analysis, the double-WD merger fraction is estimated to be about 20\% in our mass range.
\end{enumerate}
The results for the two setups are similar.

This previously unreported extra cooling delay on the Q branch is a challenge to the white dwarf cooling model and our understanding of white dwarf physics. We propose that $^{22}$Ne settling \citep{Bildsten_2001} could be the physical origin of this extra cooling delay. $^{22}$Ne settling favors C/O-core versus O/Ne-core white dwarfs, suggesting that the extra-delayed white dwarfs are also double-WD merger products, i.e., our setup 1 is correct. This idea is also supported by the concentration of DQ white dwarfs and lack of wide-separation binaries on the Q branch also support this idea. To further investigate the nature of this extra cooling delay, detailed cooling models for $m_\text{WD}>1.1\,M_\odot$ C/O white dwarfs with the $^{22}$Ne settling will be needed.

High-mass white dwarfs have been used to explore the AVR, star formation history, and white dwarf mass distribution. Given the existence of the extra cooling delay, the relevant results of those functions in the literature should be reconsidered. In future analyses of these functions, one could reduce the influence of the extra cooling delay by using only the high-mass white dwarfs above the Q branch or modeling the extra cooling delay.

\section*{Acknowledgments}

We thank the anonymous referee for providing useful suggestions to improve the quality of our draft.
We thank Pierre Bergeron for providing the synthetic colors of the revised {\it Gaia} DR2 passbands, Mar\'ia E. Camisassa for providing the cooling sequences of O/Ne white dwarfs before they became public, Silvia Toonen for providing binary population synthesis results in specific mass ranges and comments on our draft, Kareem El-Badry for providing an extended version of the {\it Gaia} wide-separation binary catalog, and Hsiang-Chih Hwang for pointing out a typo in our code.
We thank Pier-Emmanuel Tremblay for his detailed comments and criticisms, which significantly helped us to improve our draft. We thank J. J. Hermes for his detailed comments and suggestions that improved the quality of our draft. We thank Josiah Schwab, Evan Bauer, and Charles Horowitz for discussions of $^{22}$Ne settling and crystallization.
We also thank Chao Liu, Kevin Schlaufman, and Rosemary Wyse for discussions.
S.C. thanks Siyu Yao for her constant encouragement and inspiration.
J.C. would like to acknowledge his support from the National Science Foundation (NSF) through grant AST-1614933.
B.M. thanks the David and Lucile Packard Foundation.

This project was developed in part at the 2019 Santa Barbara Gaia Sprint, hosted by the Kavli Institute for Theoretical Physics at the University of California, Santa Barbara.

This work has made use of data from the European Space Agency (ESA) mission
{\it Gaia} (\url{https://www.cosmos.esa.int/gaia}), processed by the {\it Gaia}
Data Processing and Analysis Consortium (DPAC,
\url{https://www.cosmos.esa.int/web/gaia/dpac/consortium}). Funding for the DPAC
has been provided by national institutions, in particular the institutions
participating in the {\it Gaia} Multilateral Agreement.

\software{astropy package \citep{AstropyCollaboration_2013, AstropyCollaboration_2018}, corner.py \citep{corner}, emcee \citep{Foreman-Mackey_2013}, numpy \citep{oliphant2006guide},  matplotlib \citep{Hunter_2007}, SciPy \citep{Virtanen_2019}}

\begin{appendix}

\section{Parameterization of the extra cooling delay}
\label{app:extra-cooling-delay scenario}

\begin{figure}
    \centering
    \includegraphics[width=\columnwidth]{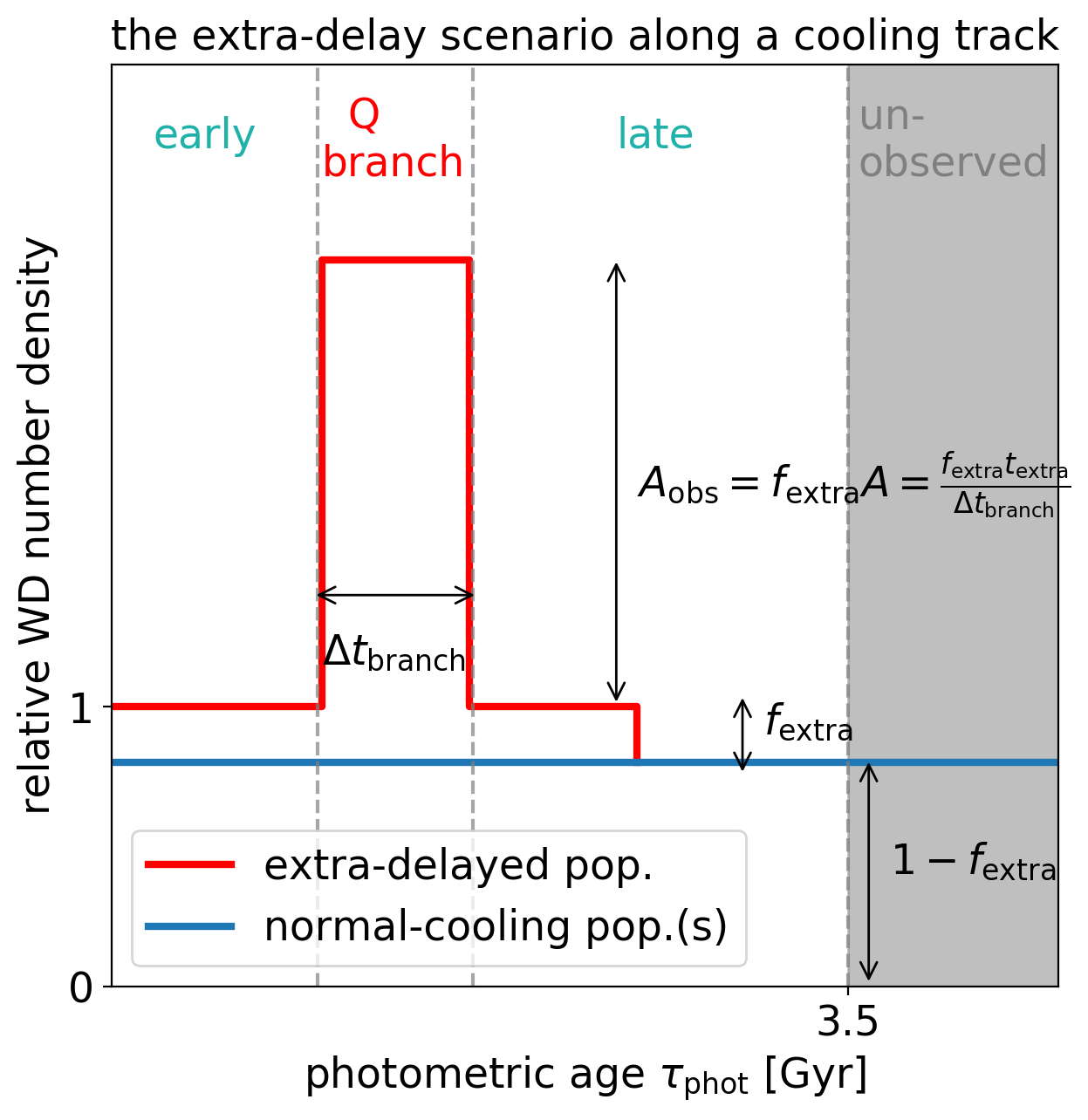}
    \caption{A sketch of the extra-delay scenario. The axes are the same as in Figure~\ref{fig:photometric age distribution}. Some WDs are subject to the extra cooling delay and the rest of them are not, corresponding to the extra-delayed and normal-cooling population in the figure. We also illustrate the quantities of equation~\ref{eq:enhancement}.\label{fig:sketch}} 
\end{figure}

Figure~\ref{fig:sketch} shows the parameterization of our extra-cooling-delay scenario on the Q branch. The observed enhancement $A_{\rm obs}$ in the photometric-age distribution can be expressed as
\begin{equation}
    \label{eq:enhancement}
    A_\text{obs}\equiv \frac{n_\text{on branch}}{n_\text{off branch}}-1=f_\text{extra}\,A = \frac{f_\text{extra}\,t_\text{extra}}{\Delta t_\text{branch}}\,,
\end{equation}
where $A$ is the intrinsic enhancement for the extra-delayed population itself. $\Delta t_\text{branch}$ is the average width of the Q branch in terms of photometric age, i.e., the time for a normal-cooling white dwarf to pass through the branch. We directly measure this width from Figure~\ref{fig:WD_HR_70} with just the photometric ages grid and our definition of the Q-branch segment, finding an average value $\Delta t_\text{branch}=$ 0.74~Gyr. White dwarfs with the extra cooling delay will spend much more time passing the branch. 

When only the enhancement $A_\text{obs}$ is used to investigate the Q branch, there is clearly a degeneracy between $f_\text{extra}$ and $t_\text{extra}$ in this two-population scenario. Velocity information helps to break this degeneracy.

\section{The Bayesian framework and elimination of selection effects}
\label{sec:bayesian framework}

We follow a Bayesian approach to build our model. This means that we can first build a forward model outputting the likelihood probability density function (PDF) of observables $\mathbfit{y}$ given the model parameters $\bm{\theta}$: 
\begin{equation}
    \mathcal{L}\equiv p(\mathbfit{y}|\bm{\theta})\,,
\end{equation}
and then we obtain the posterior PDF of model parameters $p(\bm{\theta}|\mathbfit{y})$ from the observed value of $\mathbfit{y}$ through the Bayes' theorem:
\begin{equation}
    p(\bm{\theta}|\mathbfit{y})\propto \mathcal{L}\cdot p(\bm{\theta})\,,
\end{equation}
where $p(\bm{\theta})$ is the prior PDF of the parameters. Finally, we use the MCMC method to sample the posterior distribution and estimate the parameters of interest after marginalizing nuisance parameters. Among these three steps, the key part is to construct the likelihood.

As each white dwarf provides an independent observation, the likelihood $\mathcal{L}$ in our model can be written as the product of the likelihoods of each individual white dwarf:
\begin{align}
\label{eq:likelihood}
    \mathcal{L} = \prod_i p_i(\mathbfit{y}_i| \bm{\theta})\,.
\end{align}

To avoid a direct dependence on selection effects, we use the conditional likelihood to let the constraining power originate only from velocity distributions: we define the individual likelihood $p_i$ as the probability density for the $i$th white dwarf to have transverse the velocity $\bvT$ given all other observables of this white dwarf:
\begin{equation}
\label{eq:individual likelihood}
    p_i \equiv p(\bvT|\{\tau_\text{phot},m_{\rm{WD}},l,b\}_i, \bm{\theta})\,.
\end{equation}
We condition on $\tau_\text{phot}$ and $m_{\rm{WD}}$ because their distributions are influenced by the detection completeness, quality cuts, and white dwarf spatial distribution. Moreover, the mass $m_\text{WD}$ in equation~\ref{eq:individual likelihood} model is only used to identify whether a white dwarf is on the Q branch. In order to decompose the different populations, we derive
\begin{align}
\label{eq:conditional}
    \nonumber p(\bvT|\tau_\text{phot},m_\text{WD})&=\frac{p(\bvT,\tau_\text{phot},m_\text{WD})}{p(\tau_\text{phot},m_\text{WD})}\\
    &=\frac{\sum f_x\,p_x(\bvT,\tau_\text{phot},m_\text{WD}) }{\sum f_x\,p_x(\tau_\text{phot},m_\text{WD})}\,,
\end{align}
where the sums are taken over $x$ with possible values `s', `m', and `extra' representing different populations.

To express these observable distributions by the AVR and star formation history, we employ the probability identity:

\begin{align}
    \nonumber p(\bvT,\tau_\text{phot})&=\int{p(\bvT|\tau_\text{phot},\tau)\, p(\tau_\text{phot},\tau)\,{\rm d}\tau}\\
    &=\int{p(\bvT|\tau)\,p(\tau_\text{phot},\tau)\,{\rm d}\tau}\,
\end{align}
(where the second step is valid because the velocity is only a function of the true age $\tau$) and another identity:
\begin{align}
    p(\tau_\text{phot})&=\int{p(\tau_\text{phot},\tau)\,{\rm d}\tau}\,.
\end{align}
We also assume that the age distribution is uniform, $\tau\sim U$[0, 11~Gyr]. In this way, the likelihood PDF in Equation~\ref{eq:likelihood} and \ref{eq:individual likelihood} can be expressed through the delay distributions $p(\Delta t)$ and a velocity model $p(\bvT|\tau)$.

\section{The Gaussian velocity model}
\label{app:velocity model}

Here, we describe the PDF of transverse velocities $\bvT$ and their true stellar ages $\tau$ using the AVR. The velocity distribution of disk stars in the solar neighborhood with respect to the local standard of rest can be approximated as superposition of 3D Gaussian distributions \citep[e.g.,][]{Binney_2008}:
\begin{equation}
\label{eq:3d-gaussian}
    p(\mathbfit{v}|\tau)=\frac{\exp{[-\frac{1}{2}(\mathbfit{v}-\mathbfit{v}_0)^T\mathbf{\Sigma}(\tau)^{-1}(\mathbfit{v}-\mathbfit{v}_0)]}}{\sqrt{8\pi^3|\bm{\Sigma}(\tau)|}}\,,
\end{equation}
whose mean and covariance matrix are determined by stellar age $\tau$. The mean velocity $\bvzero(\tau)$ is determined by two effects: the solar reflex motion ($-U_{\odot}$, $-V_{\odot}$, $-W_{\odot}$) with respect to the local standard of rest, and the asymmetric drift in $V$ direction by $-\sigma_{\rm U}^2/80\rm\,km\,s^{-1}$ \citep[e.g.,][]{Binney_2008}. We set the solar motion as free parameters and use them to check the validity of our model.

To obtain the distribution of the observable transverse velocity $\bvT=(\vL,\vB)^T$, we project the 3D Gaussian distribution $p(\mathbfit{v}|\tau)$ onto the tangential plane for each white dwarf and marginalize the radial component $v_{\rm R}$. Because the resulting distribution is still a Gaussian distribution for a given age $\tau$, the only task is to find its covariance matrix and mean vector. Let $\bm{v}_{\rm XYZ}=(U,V,W)$ and $\bm{v}_{\rm \rm LBR}=(v_{\rm L},v_{\rm B},v_{\rm R})$ be the expressions of the same vector $\bm{v}$ in the XYZ and LBR coordinate systems, respectively, and matrix $M$ the rotation transformation matrix between the two systems:
\begin{equation}
\label{eq:rotate}
    (\bm{v}-\bm{v}_{0})_{\rm XYZ}=M\cdot (\bm{v}-\bm{v}_{0})_{\rm LBR}\,,
\end{equation}
where
\begin{equation}
    M=\begin{bmatrix} 
    \sin l&-\sin b\cos l&\cos b \cos l\\
    \cos l&-\sin b \sin l&\cos b \sin l\\
    0&\cos b&\sin b
    \end{bmatrix}\,.
\end{equation}
We ignore the small in-disk rotation between the Cartesian coordinate $XYZ$ and the galactic polar coordinate. Then, we write the 3D Gaussian distribution in both coordinate systems (note that the Jacobian determinant of rotation transform is unity) and obtain the following relation:
\begin{equation}
    (\bm{v}-\bm{v}_{0})_{\rm XYZ}^T\mathbf{\Sigma}_{\rm XYZ}^{-1}(\bm{v}-\bm{v}_{0})_{\rm XYZ}=(\bm{v}-\bm{v}_{0})_{\rm LBR}^T\mathbf{\Sigma}_{\rm LBR}^{-1}(\bm{v}-\bm{v}_{0})_{\rm LBR}\,.
\end{equation}
Substituting equation~\ref{eq:rotate}, we obtain:
\begin{equation}
\label{eq:rotate 3d-gaussian}
    \mathbf{\Sigma}_{\rm LBR}=(M^T\mathbf{\Sigma}_{\rm XYZ}^{-1}M)^{-1}=M^T\mathbf{\Sigma}_{\rm XYZ}M\,,
\end{equation}
where we assume
\begin{equation}
\mathbf{\Sigma}_{\rm XYZ}=\begin{bmatrix}
    \sigma_{\rm U}^2&0&0\\
    0&\sigma_{\rm V}^2 & 0\\
    0&0&\sigma_{\rm W}^2
\end{bmatrix}\,.
\end{equation}
The covariance matrix $\mathbf{\Sigma}_{\rm LB}$ for the 2D Gaussian distribution marginalized along the $R$ direction is the top left 2$\times$ 2 sub-matrix of $\mathbf{\Sigma}_{\rm LBR}$. The mean vector $\mathbfit{v}_{\rm T0}=(\bm{v}_{0})_{\rm LB}$ can be derived directly from vector rotation and projection. Then, the conditional PDF of the two transverse components $\bvT=(\vL,\vB)$ of $\bm{v}$ can be written as
\begin{equation}
\label{eq:2d-gaussian}
    p(\bvT|\tau,l,b)=\frac{\exp{[-\frac{1}{2}(\bvT - \mathbfit{v}_{\rm T0})^T\bm{\Sigma}_{\rm LB}^{-1} (\bvT - \mathbfit{v}_{\rm T0})]}}{\sqrt{4\pi^2|\bm{\Sigma}_{\rm LB}|}},
\end{equation}
where the covariance matrix $\bm{\Sigma}_{\rm LB}(\tau,l,b)$ and mean $\mathbfit{v}_{\rm T0}(\tau,l,b)$ depends on $(\tau,l,b)$. We use the condition on $(l,b)$ for each white dwarf to avoid modeling the spatial selection effects and reduce unnecessary parameters and biases. 

\section{MCMC Settings}
\label{app:MCMC}

\begin{figure*}
    \centering
    \includegraphics[width=\textwidth]{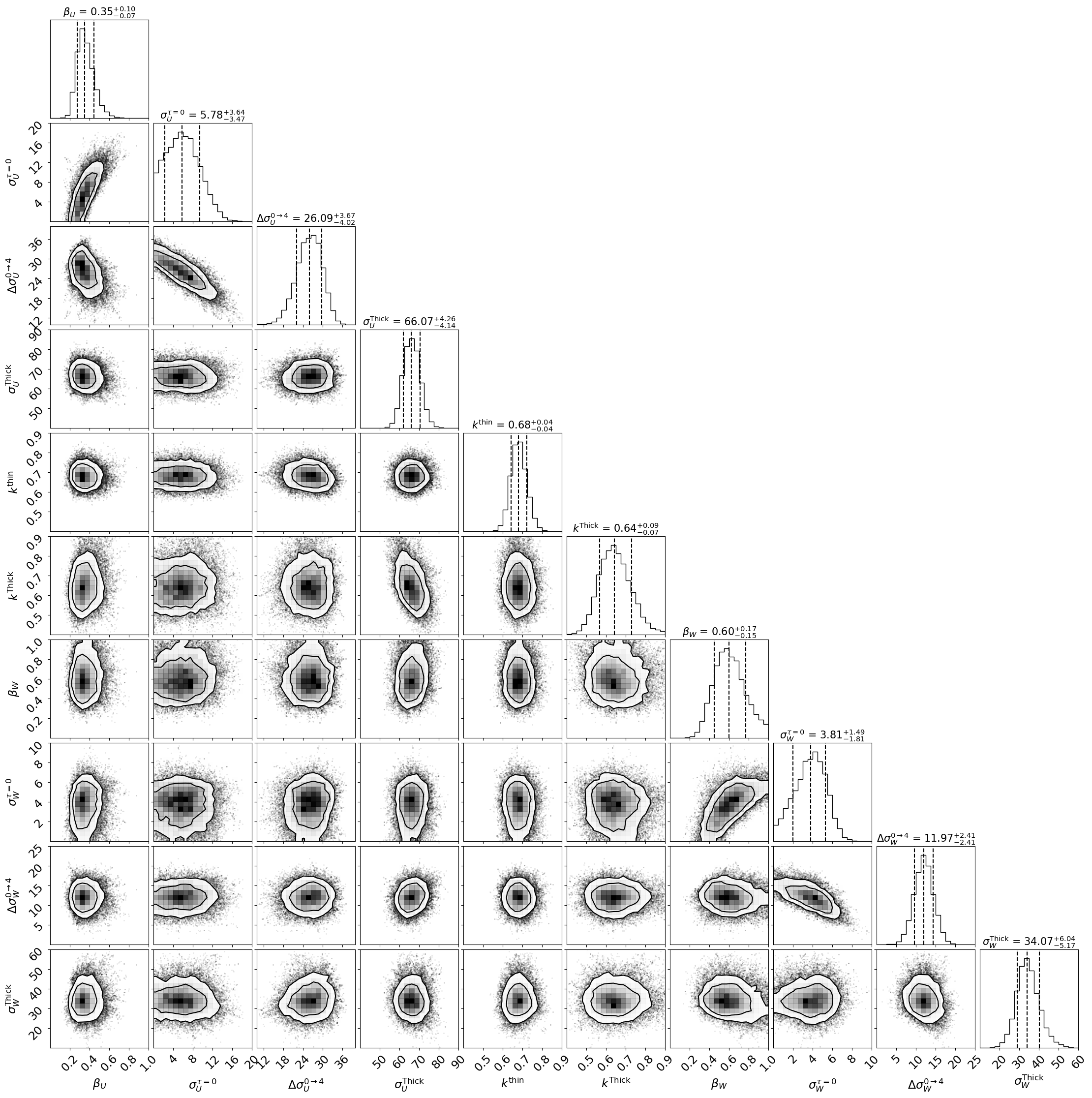}
    \caption{\label{fig:MCMC_app}The corner plot of the posteriors of the AVR parameters. We use flat priors for these parameters within the ranges shown on this figure. We have checked that there are no correlations between these parameters and the three main parameters $f_\text{extra}$, $t_\text{extra}$, and $f_\text{m}$, and the three components of solar motion in our model.}
\end{figure*}

In our Bayesian model, we assume uniform distributions for parameter priors. We feed the affine invariant MCMC sampler \texttt{emcee} \citep{Foreman-Mackey_2013} with the natural logarithm of the likelihood function defined in equation~\ref{eq:likelihood}. We use 500 walkers to explore the parameter space. After 200 steps of burn-in, the chains are checked to converge by comparing the percentile values of the parameters in each chain. Then, we run another 400 steps and use this sampling to represent the posterior distribution of each parameter. Figure~\ref{fig:MCMC_app} shows the marginal posteriors of the 10 parameters of the AVR and their correlations, under the setup 1. Setup 2 leads to similar constraints of the AVR.

\section[The peak of 22Ne-settling effect]{The peak of $^{22}$Ne-settling effect}
\label{app:22Ne}

The delay effect depends on the fractional contribution $L_\text{extra}/L_\text{surf}$ of the extra source luminosity to the surface luminosity of the white dwarf because the more this extra energy contributes, the less does the white dwarf need to consume its thermal energy and to cool down. The pile-up factor of this effect can be expressed as
\begin{align}
\label{eq:22Ne}
    A = \frac{\zeta^{-1}}{\zeta_0^{-1}}-1 = \frac{1}{L_\text{surf}/L_\text{extra}-1}\,,
\end{align}
where $A$ is the same as defined in Equation~\ref{eq:enhancement}, $\zeta$ and $\zeta_0$ are the cooling rates with and without the extra energy. Assuming no crystallization suppression, $L^\text{Ne}_\text{extra}$ will meet $L_\text{surf}$ at some surface temperature $T_\text{eff}$. If this occurs, $^{22}$Ne settling will stop the white dwarf cooling at this temperature until $^{22}$Ne is exhausted, creating a peaked delay effect. Here, we estimate the dependence of this meeting temperature as a function of white dwarf mass, which can be translated into a curve on the H--R diagram. 

According to \citet{Bildsten_2001}, the energy release of $^{22}$Ne settling can be expressed by
\begin{align}
\label{eq:22Ne power}
    L^\text{Ne}_\text{extra}=\int_0^R FVn(^{22}\text{Ne})4\pi r^2 dr\,,
\end{align}
where $F=2m_{\rm p} g_r$ is the net force felt by each nucleus of $^{22}$Ne, $g_r$ is the gravity at radius $r$; $V=(D/D_s)18m_p g_r/(Ze\Gamma^{1/3}\sqrt{4\pi \rho})$ is the drift velocity of the settling, $\Gamma\equiv (Ze)^{2}/(akT)\propto \rho^{1/3}Z^2/(TA^{1/3})$ is the Coulomb coupling parameter, $D$ is the inter-diffusion coefficient of $^{22}$Ne, and $D_s$ is the one-component self-diffusion coefficient, which can be used as a reference value. Substituting these quantities in Equation~\ref{eq:22Ne power}, we obtain:
\begin{align}
    L^\text{Ne}_\text{extra}&\propto \int_0^R g_r \cdot \frac{D}{D_s} X(^{22}{\rm Ne}) \frac{A^{0.11}}{Z^{1.67}} g_r\rho^{-0.61} T_c^{0.33} \cdot \rho r^2 dr\,,
\end{align}
where $X$ is the element abundance in mass, $T_c$ is the core temperature. For the main composition of a white dwarf, the charge-to-mass ratio $Z/A=0.5$ is a constant. Assuming the following approximations: $g_r \sim g \sim M/R^2$, $\rho \sim g/R$, $\int_0^R dr \sim R$, we obtain
\begin{align}
    L^\text{Ne}_\text{extra}
    &\propto \frac{D}{D_s} X(^{22}{\rm Ne}) Z^{-1.56} T_c^{0.33} g^{2.39} R^{0.61} \cdot R^2
\end{align}
Before the convective coupling between the core and atmosphere, the core temperature $T_c$ scales with the surface temperature $T_\text{eff}$ and gravity $g$ as
\begin{align}
\label{eq:Teff}
    T_c &\propto (T_\text{eff}^4/g)^{0.41}\,,
\end{align}
which is obtained empirically from existing white dwarf models \citep{Fontaine_2001}. This scaling relation is more realistic than the one used by \citet{Mestel_1952}. Substituting Equation~\ref{eq:Teff} for $T_c$, we obtain
\begin{align}
    L^\text{Ne}_\text{extra}
    \propto \frac{D}{D_s} X(^{22}{\rm Ne}) Z^{-1.56} T_\text{eff}^{0.55} g^{2.25} R^{0.61} \cdot R^2\,.
\end{align}
We checked the simulation results from Figures 7 and 8 of \citet{Garcia-Berro_2008} and found that our scaling relation is accurate to within 10\%, which is sufficient for our purpose.

When the surface luminosity $L_\text{surf} \propto T_\text{eff}^4\cdot R^2$ is set equal to $L^\text{Ne}_\text{extra}$, we obtain
\begin{align}
    T_\text{eff} &= K^{0.29} g^{0.65} R^{0.18}\,,
\end{align}
where $K \propto \frac{D}{D_s} X(^{22}{\rm Ne}) Z^{-1.56}$ is a constant that has no relevance to the white dwarf mass $M$, and both $g$ and $R$ are determined by $M$. The proportional factor within $K$ can be evaluated from an existing simulation of $^{22}$Ne settling.

\end{appendix}

\bibliography{Qbranch}
\end{CJK*}
\end{document}